\documentclass[aps,10pt,prd,preprintnumbers,superscriptaddress,twocolumn,amsmath,amssymb,nofootinbib,a4paper,showpacs]{revtex4-1}

\usepackage{amsfonts,amsthm}
\usepackage{bm,bbm}
\usepackage{graphicx}
\usepackage[T1]{fontenc}
\usepackage{color}
\usepackage[breaklinks]{hyperref}
\usepackage[titletoc]{appendix}

\newcommand{\be}{\begin{equation}}
\newcommand{\ee}{\end{equation}}
\newcommand{\ba}{\begin{eqnarray}}
\newcommand{\ea}{\end{eqnarray}}
\def\bs{\begin{subequations}}
\def\es{\end{subequations}}
\def\a{\alpha}
\def\b{\beta}
\def\de{\delta}
\def\g{\gamma}
\def\la{\lambda}

\def\e{\epsilon}

\def\om{\omega}
\def\G{\Gamma}

\def\s{\sigma}
\def\vr{\varrho}

\def\N{\nabla}

\def\cC{\mathcal{C}}
\def\cD{\mathcal{D}}
\def\cE{\mathcal{E}}

\def\cG{\mathcal{G}}

\def\cK{\mathcal{K}}
\def\cL{\mathcal{L}}
\def\cM{\mathcal{M}}

\def\cP{\mathcal{P}}

\def\cV{\mathcal{V}}
\def\bE{\mathbbm{e}}

\def\Or{O}
\def\ds{d_{\rm S}}
\def\dh{d_{\rm H}}
\def\dw{d_{\rm W}}

\def\p{\partial}

\def\B{\Box}
\newcommand{\Eq}[1]{(\ref{#1})}
\def\com{\color{magenta}}
\def\cob{\color{blue}}

\def\llangle{\langle\!\langle}
\def\rrangle{\rangle\!\rangle}
\newcommand{\book}[5]{\emph{#1}  (#2, #3, #4, #5)}
\newcommand{\books}[4]{\emph{#1} (#2, #3, #4)}

\newcommand{\arX}[1]{\href{http://arxiv.org/abs/#1}{{\ttfamily\com arXiv:#1}}}
\newcommand{\doin}[6]{\href{http://dx.doi.org/#1}{{\cob #2 #3 {\bf #4}, #5 (#6)}}}
\newcommand{\doinn}[5]{\href{http://dx.doi.org/#1}{{\cob #2 {\bf #3}, #4 (#5)}}}
\newcommand{\doij}[5]{\href{http://dx.doi.org/#1}{{\cob #2 #3 (#5) #4}}}

\newcommand{\tia}[1]{}
\newcommand{\boxd}[1]{\boxed{\phantom{\Biggl(}#1\phantom{\Biggl)}}}

\def\rme{e}
\def\rmd{d}
\def\rmi{i}

\begin{document}

\title{Spectral dimension and diffusion in multiscale spacetimes}

\author{Gianluca Calcagni}
\email{calcagni@iem.cfmac.csic.es}
\affiliation{Instituto de Estructura de la Materia, CSIC, Serrano 121, 28006 Madrid, Spain}

\author{Giuseppe Nardelli}
\email{nardelli@dmf.unicatt.it}
\affiliation{Dipartimento di Matematica e Fisica, Universit\`a Cattolica, via Musei 41, 25121 Brescia, Italy}
\affiliation{INFN Gruppo Collegato di Trento, Universit\`a di Trento, 38100 Povo (Trento), Italy}

\begin{abstract}
Starting from a classical-mechanics stochastic model encoded in a Langevin equation, we derive the natural diffusion equation associated with three classes of multiscale spacetimes (with weighted, ordinary, and ``$q$-Poincar\'e'' symmetries). As a consistency check, the same result is obtained by inspecting the propagation of a quantum-mechanical particle in a disordered environment. The solution of the diffusion equation displays a time-dependent diffusion coefficient and represents a probabilistic process, classified according to the statistics of the noise in the Langevin equation. We thus illustrate, also with pictorial aids, how spacetime geometries can be more completely catalogued not only through their Hausdorff and spectral dimension, but also by a stochastic process. The spectral dimension of multifractional spacetimes is then computed and compared with what was found in previous studies, where a diffusion equation with some open issues was assumed rather than derived. These issues are here discussed and solved, and they point towards the model with $q$-Poincar\'e symmetries.
\end{abstract}

\date{April 9th, 2013}


\preprint{\doin{10.1103/PhysRevD.88.124025}{Phys.\ Rev.}{D}{88}{124025}{2013} [\arX{1304.2709}]}

\maketitle


\tableofcontents

\section{Introduction}

The concept of \emph{spectral} (or \emph{fracton}) \emph{dimension} $\ds$ of spacetime descends from the analogous one for a set, be it a fractal or a Riemannian manifold \cite{AO,RT,Wat85,HBA,KiL,bH}. The idea is to have a test particle diffuse in the set (a nonrelativistic ``space''), and see how it behaves. This process is governed by a diffusion equation, which for a Brownian motion reads
\be\label{diet}
(\p_t-\kappa_1\N^2_x)P(x,x',t)=0\,,\quad P(x,x',0)=\de(x-x')\,,
\ee
where the first-order derivative $\p_t$ is the diffusion operator in time $t$; $\kappa_1$ is a constant; $\N^2_x$ (called spatial generator in probability theory) is the natural Laplacian in the given space (eventually containing metric structure) acting on the $x$ dependence of the solution $P$; and $x'$ is the initial point where the probe, a pointwise particle represented by the $\de$ initial condition at $t=0$, starts diffusing. The process has a probabilistic interpretation when $P$ is a probability density function (PDF), $P\geq 0$ for all $x$. In particular, the random position field $X(t)$ associated with Eq.\ \Eq{diet} is a Brownian motion. The scaling of the variance (i.e., how the mean-squared displacement $\langle X^2\rangle$ of the process $X$ increases in time) is related to $\ds$.

To make sense of this picture in a spacetime context with $D$ topological dimensions (one timelike and $D-1$ spacelike), one must make a few strong assumptions on this diffusive process:
\begin{enumerate}
\item[(A)] Replace the Laplacian $\N^2$ with the covariant Laplace--Beltrami (or d'Alembertian) operator $\B=\N_\mu\N^\mu$ in imaginary time. Thus, there are $D$ directions parametrized by coordinates $x^\mu$, $\mu=1,\dots, D$, where $x^D=\rmi t$ and $t=x^0$ is time in Lorentzian signature. In order to keep the probabilistic interpretation of the diffusion equation, the spatial generator is assumed to be an elliptic operator, hence the requirement of Euclideanization.
\item[(B)] Time is thus treated on equal footing with the other coordinates, in conformity with the spirit of general relativity. Consequently, the variable $t$ in the diffusion equation \Eq{diet} is replaced by an abstract evolution parameter $\s$ everywhere.
\item[(C)] In particular, $\s$ has the same dimensionality as $t$ (it is a length or time scale, and $[\s]=-1$ in momentum units), and the diffusion operator $\p_t$ in \Eq{diet} is simply replaced by $\p_\s$.
\end{enumerate}
This construction may be unsatisfactory for various reasons. In general-relativistic systems, the diffusion equation is expected to have a covariant form, which should survive even under the simple replacement procedure (B)--(C). Yet, a nonrelativistic diffusion equation is typically assumed for curved manifolds. Second, statistical-mechanics time $t$ is Euclideanized in the spatial generator $\B$, but not in the diffusion operator (which would result in a Schr\"odinger equation). The general feeling is that, even when the diffusion equation is more or less well motivated in the context of statistical mechanics, it carries a considerable level of arbitrariness when promoted to the diffusion equation associated with a given spacetime. In this sense, the diffusion equation (and, hence, the spectral dimension $\ds$) of a spacetime cannot be truly derived from solid first principles. 

For ordinary spacetimes, the proof of the pudding is in the eating, and the results of the procedure (A)--(C) are reasonable enough not to require a revision. We can better appreciate the problem, however, when moving to \emph{multiscale spacetimes}, in particular those with multifractional measure \cite{fra4,frc1,frc2,frc3,ACOS,fra6,frc4,frc5,fra7,frc6}. These spacetimes \cite{fra4,AIP} have been recently introduced as realizations of geometries with anomalous (or even fractal-like) properties \cite{fra1,fra2,fra3}, which appear in certain regimes of many quantum-gravity models (a canonical list of references can be found, e.g., in \cite{frc2,frc4,fra1,Car09,SVW2}). The geometry of multiscale spacetimes is continuous and tagged by $D$ topological dimensions, but spacetime points $x$ contribute with different weights to the Lebesgue measure $\rmd^D x\,v(x)$. A hierarchy of scales characterizing the geometry is included in the weight $v(x)$, which is chosen by striking a compromise between technical feasibility (analytic progress is difficult in the most general situation \cite{fra1,fra2,fra3}) and the realization, still in a rigorous manner, of anomalous scaling and symmetry properties of irregular and multifractal geometries. Apart from developing an autonomous phenomenology in particle physics and quantum gravity with a number of interesting properties, this framework helped to clarify how to use various concepts of transport theory, complex systems and fractal geometry to gain physical insight into quantum-spacetime models such as noncommutative spacetimes \cite{ACOS}, asymptotic safety and Ho\v{r}ava--Lifshitz gravity \cite{fra7,CES}.

The content of the present paper can be summarized in seven points.
\begin{itemize}
\item[(a)] \emph{Addressing issues of previous formulations}. The spectral dimension of multifractional spacetimes has been computed in \cite{frc1,frc6,fra2} using a certain ansatz for the diffusion equation, and found to be anomalous ($\ds\neq D$). Such diffusion equation, however, was simply assumed and it turns out to have some issues, which we discuss in Sec.\ \ref{old}. We completely revisit the problem (and address these issues) from the more fundamental point of view of statistical and quantum mechanics. Although the resulting diffusion equation will be different from the old assumption, the spectral dimension will still be anomalous, i.e., different from the integer topological dimension $D$. 
\item[(b)] \emph{Analytic control of multiscale geometries}. While in several quantum-gravity models the spectral dimension of \emph{multiscale} geometries is quoted in terms of asymptotic regimes, here we have full analytic control of the whole dimensional flow and we can follow the behaviour of the spectral dimension at any given scale. The advantages include having clear scale identifications, a precise separation between ultraviolet and infrared regimes, an exact treatment of transition regimes, and so on.
\item[(c)] \emph{Fractal versus nonfractal geometries}. One must impose certain conditions in order for multifractional spacetimes to be fractal in the usual sense. In practice, only one multiscale model (with so-called $q$-Poincar\'e symmetries) will satisfy all these requisites. Such a clarification of terminology, perhaps obvious for a mathematician, has a wider scope than multiscale models: it states conditions by which any model of quantum geometry must abide in order to be labeled as ``fractal.'' There is a quantitative consequence one can draw from this apparently academic discrimination: any quantum-gravity theory with a genuine fractal geometry would predict, in general, a specific density of energy states $\rho(E)$ [Eq.\ \Eq{rhoed}]. This density of states can be interpreted as that associated with modes of virtual particles in vacuum. It can change in the presence of matter or in extreme spacetime configurations, where energy levels get populated. For instance, near a black hole there is a special density of states reproducing the entropy area law \cite{Pad98,Pad99}, which drastically modifies the diffusion properties of spacetime \cite{AC1}. However, the vacuum structure is fixed once and for all, and this is important precisely to link simple regimes of the theory with such extreme limits. Here we do not develop this topic in detail but we mention it as one of the possible applications of our results to quantum gravity.
\item[(d)] \emph{Role of statistical mechanics}. With respect to previous studies, the role of statistical mechanics in the determination of the diffusion equation is absolutely dominant and more powerful than any \emph{ad hoc} prescriptions for the diffusion equation. This is a novelty in the field of quantum gravity.
\item[(e)] \emph{Degeneracy problem and role of stochastic processes}. As in \cite{fra6,frc4,CES}, one of the key goals of the present work is to show that the spectral dimension is only one of the many possible ways to characterize a spacetime geometry: it is well known that an anomalous correlation function, whose exponent is governed by $\ds$, can be obtained by a number of inequivalent diffusion processes \cite{Sok12}. A stronger version of the problem points out that some processes may even have the same diffusion equation; the example of scaled versus fractional Brownian motion is typical \cite{CES,Sok12}. Many quantum-gravity theories share the same spectral dimension but they may be not physically equivalent for these reasons. To remove the degeneracy, it is important to gain insight into the type of stochastic process underlying the diffusion equation. Thus, \emph{spacetime geometries can be classified not only according to their Hausdorff and spectral dimension, but also by a stochastic process}. More generally, there exists a well-furbished ``alternative toolbox'' \cite{fra4,frc4,fra7} mutuated from various disciplines of mathematics and mathematical physics (chaos theory, probability theory, transport and percolation theory, statistical mechanics, complex systems, multifractal geometry) which is still to be fully tapped into by the quantum-gravity community, where discussions are often limited only to a few geometric indicators (Hausdorff and spectral dimension).
\item[(f)] \emph{Graphic examples of point (e)}. We will provide a visual representation of the possibilities of this arsenal by plotting the trajectories of random walkers associated with different fractional geometries. The figures in the text illustrate how to discriminate among anomalous geometries via the information provided by the stochastic process. 
In principle, one could make measurements of a ``Brownian'' particle in an anomalous spacetime and check that its trajectory does not actually possess the characteristics of Brownian motion. A detailed analysis can identify the geometry of space uniquely---for instance comparing each of the three stochastic processes \Eq{fsbm}, \Eq{scabm}, and \Eq{fsbmq} of the figures, corresponding to three inequivalent multiscale spacetimes, with the reference trajectory of ordinary Brownian motion. Of course, at molecular scales spacetime is ordinary and one should go at much higher energies to probe multiscale effects. Still, the main message is clear: spacetimes with same or similar spectral dimensions can be discriminated on fairly physical grounds by methods more refined than those conventionally used in other theoretical physics contexts.
\item[(g)] \emph{Adjointness problem and role of stochastic quantum mechanics}. Other stimulating results come from this study. {\it (g1)} In transport theory, it is common to find diffusion equations in which the spatial generator is not self-adjoint with respect to the natural scalar product of the Hilbert space on which it acts. In practice, this amounts to an adjointness condition on the Laplacian. This is not an issue by itself, since the diffusion equation is determined by the physics of the problem and one simply abides with it. However, diffusion equations of spacetime are a completely different matter: they determine the spectral dimension and other indicators of the underlying geometry, and one may worry about cases where the transport equation is not adjoint. This situation can happen quite generally in quantum gravity, as soon as one considers effective geometries where the scalar-product structure is deformed. Then, it is important to ask whether the given diffusion equation and its adjoint counterpart yield the same spectral dimension. If not, one might cast serious doubts about the whole procedure. We may call this the ``adjointness problem.'' To put it differently, suppose two models of anomalous geometry happen to sport mutually adjoint diffusion equations. (This is the present case of multiscale theories with, respectively, ordinary and weighted Laplacians.) Physically, the two systems are different, since they are associated with different PDFs solving different diffusion equations. However, they may fall into the same class of geometries labeled by a given spectral dimension $\ds$. Most of the time, the assumed diffusion equation is self-adjoint (as, for instance, in asymptotic safety and Ho\v{r}ava--Lifshitz gravity \cite{fra7,CES}), but this easy situation can also induce one to take quantum-gravity diffusion equations acritically. We propose the adjointness problem for the first time, providing a concrete example of two models where the diffusion equations are mutually adjoint. On this testing ground, we can make a precise clarification of the problem. In Secs.\ \ref{ordi} and \ref{muso}, we show that the two models yield the same spectral dimension, and that therefore they do fall into the same class of geometries. On our way to this result, we highlight several interesting mathematical and physical subtleties. {\it (g2)} The probability density is recognized as a bilinear functional, grounded on a stochastic limit of quantum mechanics. Also this aspect is, to the best of our knowledge, new in quantum gravity and has a twofold impact. First, it provides an independent check on the assumptions underlying the structure of the Langevin equation, an object which is more fundamental than the diffusion equation but, quite often, not derivable from first principles in a solid way. In particular, there is a clear, direct relation between diffusion time and quantum-mechanical time. Second, it offers a novel basis from which to construct the diffusion equation from first principles. Given a theory of quantum gravity, if it was possible to study the quantum mechanics of a particle on the effective spacetime of the theory, then one would have a means to derive the stochastic diffusion equation from it, following the same steps as in ordinary transport theory. This possibility may be contemplated, for instance, in asymptotic safety and help to identify uniquely the diffusion equation there \cite{CES}.
\end{itemize}
The plan of the paper is as follows. In Sec.\ \ref{euc}, we review the case of ordinary Minkowski spacetime, starting from the derivation of the diffusion equation under the assumption of having a random walker with a certain stochastic interpretation (Brownian motion). This standard setting will introduce the main tools of statistical mechanics and probability and transport theory needed also in multiscale spacetimes, the latter reviewed in Sec.\ \ref{mfs}. The reader familiar with both diffusion theory and multiscale spacetimes can safely skip these parts, including Sec.\ \ref{lapl}. In Sec.\ \ref{old}, we analyze the diffusion equation previously assumed for these spacetimes and its related problems. Section \ref{diffs} presents the new results for various classes of fractional spacetimes, using and extending the alternative toolbox introduced in Sec.\ \ref{euc}. The spectral dimension of multiscale geometries is computed in Sec.\ \ref{mus}, where it is shown that the theory with weighted Laplacian can have an ultraviolet regime $\ds\sim 0$ provided one interprets the scale in the measure as one signaling a transition between a ``fuzzy'' and a continuous regime. Section \ref{disc} is devoted to discussion. Appendices \ref{qmbil} and \ref{qmtime} contain digressions on how some of the results related to anomalous scaling can be obtained independently (and in agreement with the methods of the main text) from quantum mechanics.


\section{Brownian motion, diffusion equation and spectral dimension}\label{euc}

The spectral dimension of a smooth classical manifold is an indicator of the geometry and topology of spacetime. It is obtained by letting a test particle diffuse on the Euclidean version of the manifold, and calculating the probability to find the probe again at the starting point after some diffusion time $\s$ (not to be confused with coordinate time $t$, which is treated on the same grounds as a spatial coordinate). This return probability is a function $Z(\s)$ of diffusion time; if $Z\sim \s^{-\ds/2}$ is a power law, then the spectral dimension is the exponent  $\ds$.

The test particle is assumed to follow a random walk of Brownian type. This is because the associated probability density function yields, in the absence of curvature, the correct spectral dimension $\ds=D$, coinciding with the topological dimension of spacetime. The diffusion equation of Brownian motion \cite{Fic55,Ein05,Smo06} (see \cite{KTH,Zwa01} for extensive presentations) can be derived from a Langevin equation. We review Brownian motion and its diffusion equation in a nonrelativistic space in Secs.\ \ref{brmo} and \ref{22}. Promotion to the spacetime picture according to (A)--(C) will take place in Sec.\ \ref{stsd}.


\subsection{Brownian motion}\label{brmo}

Let $X(t)$ be the random variable denoting the position of a particle at time $t$; in $D$ dimensions, it is a vector. A Wiener process (or Brownian motion) $X_{\textsc{bm}}(t)$ with initial condition $X_{\textsc{bm}}(0)=x'$ is such that
\begin{enumerate}
\item[(i)] $X_{\textsc{bm}}$ is continuous in $t$ almost surely (i.e., with probability 1);
\item[(ii)] the increments of $X_{\textsc{bm}}$ are uncorrelated, meaning that $X_{\textsc{bm}}(t_2)-X_{\textsc{bm}}(t_1)$ is independent of $X_{\textsc{bm}}(t_4)-X_{\textsc{bm}}(t_3)$ if the intervals $(t_1,t_2)$ and $(t_3,t_4)$ do not overlap, $(t_1,t_2)\cap(t_3,t_4)=\emptyset$;
\item[(iii)] $X_{\textsc{bm}}$ is governed by a Gaussian distribution.
\end{enumerate}
To make the last requirement explicit, let us introduce a dimensionless white noise $\eta(t)$, that is, a Gaussian random field such that
\be\label{wn}
\langle\eta(t)\rangle_\eta=0\,,\qquad \langle\eta(t)\eta(t')\rangle_\eta=\kappa_1\de(t-t')\,,
\ee
where $\kappa_1$ is a constant with engineering dimension $[\kappa_1]=-1$. The noise is called ``white'' because its spectrum (the Fourier transform of its two-point correlation function) is a constant independent of the frequency. Here, angular brackets with subscript $\eta$ denote the average over the stochastic background, i.e., with respect to the PDF $u(\eta,t)$ of the process: for any $f$, $\langle f(\eta,t)\rangle_\eta:=\int \rmd^D \eta\, u(\eta,t) f(\eta,t)$. In the absence of external forces, the \emph{Langevin equation} of Brownian motion is
\be\label{gle0}
\p_t X_{\textsc{bm}}(t) = \eta(t)\,.
\ee
By definition, it represents a Wiener process as the integral of a white noise:
\be\label{glei}
X_{\textsc{bm}}(t) := x'+\int_0^t \rmd t'\,\eta(t')\,.
\ee
This way, the differential $\rmd X_{\textsc{bm}}$ is well defined even if the trajectory $X_{\textsc{bm}}(t)$ is nowhere differentiable in the ordinary sense. Two key features of Brownian motion are that it obeys the scaling property
\be\label{sca0}
X_{\textsc{bm}}(\la t)=\la^{\frac{1}2} X_{\textsc{bm}}( t)\,,
\ee
and that it possesses stationary increments, i.e., 
\be\label{incre}
\langle [X(t)-X( t')]^2\rangle= \langle X^2(t- t')\rangle\,.
\ee
Here, angular brackets without subscript denote the average with respect to the PDF $\hat P$ of the process, $\langle f(X, t)\rangle:=\int \rmd^D X\, \hat P(X,x', t) f(X, t)$.


\subsection{Diffusion equation}\label{22}

A Fokker--Planck equation (namely, the diffusion equation) for the PDF of Brownian motion can be derived starting from the {Langevin equation} \cite{KTH,Zwa01}
\be\label{le0}
m \ddot X+m\g\dot X+U'(X)=F\,,
\ee
where $[m]=1=[\g]$, $[F]=2$, and dots and primes denote derivatives with respect to, respectively, time $t$ and $X$. This equation describes the motion of a particle $X(t)$ of mass $m$ in the presence of a friction force (proportional to the mass of the particle, a constant coefficient $\g$ and the velocity $V=\dot X$), an external potential $U$, and a random force $F$, representing the pushing around of the particle by the medium. A writing of Eq.\ \Eq{le0} where the left- and right-hand sides are dimensionless is
\be\label{le00}
\frac{1}{\g} \ddot X+\dot X+\frac{U'(X)}{m\g}=\eta\,,\qquad \eta:=\frac{F}{m\g}\,.
\ee
There are various methods to find the diffusion equation. One takes the Langevin Eq.\ \Eq{le00} exactly and solves it. Consider for instance the case without potential, $U=0$. Integrating twice the Langevin equation $\dot V+\g V=\g\eta$ starting from time $t=0$, one obtains
\ba
X(t) &=&\int_{0}^t\rmd t' V(t')\nonumber\\
      &=& x'+ V_0\chi(t)+\g\int_{0}^t\rmd t''\,\chi(t-t'')\eta(t'')\,,\label{xgen}
\ea
where $V_0=V(0)$ is the initial velocity 
 and
\be
\chi(t-t''):=\frac{1-\rme^{-\g(t-t'')}}{\g}\,.
\ee
Expression \Eq{xgen} holds true also in more general situations with a more complicated function $\chi$, for instance when $U\neq 0$, when the friction term $m\g\dot X$ in \Eq{le0} is replaced by a nonlocal operator and the process is non-Markovian, and for general colored noise $\eta$. If the stochastic properties of $\eta$ are known, the PDFs resulting from these generalized Langevin equations (GLEs) \cite{Zwa01,WaL,Wan92,MaW,WaT,Lut01,LiM02,MeK04,JeM} can be found by elegant techniques, for instance using the characteristic (or generating) functional approach \cite{KTH,Wan92,WaT}, the functional derivative approach \cite{MaW}, or the conservation-law approach \cite{Zwa01}. In particular, the diffusion equation for Brownian motion (also known as the Smoluchowski equation when $U=0$) is obtained from Eq.\ \Eq{xgen} under three assumptions: (i) that $\eta$ is a Gaussian white noise, Eq.\ \Eq{wn} (thus, $X$ and the velocity field $V$ are also Gaussian); (ii) that the limit $V_0\to 0$ is taken (Maxwell distribution of velocities); (iii) that the times are much larger than the relaxation time $t_{\rm rt}=1/\g$. The latter corresponds to taking $\chi\approx {\rm const}$ from the start. Namely, setting $U=0$ and dropping the second-derivative term in \Eq{le00}, we have $\dot X\approx \eta$, which is nothing but \Eq{gle0}.

In fact, for our purposes it suffices to apply the conservation-law approach \cite{Zwa01}, valid for Markovian systems, and to make direct use of Eq.\ \Eq{gle0}. Although the following steps are well known, we report them to illustrate the relation between Langevin and diffusion equation for later use. The starting point is the conservation of the probability density,
\be\label{hatu1}
\langle 1\rangle=\int \rmd^DX\, \hat P(X,x',t) =1\qquad \forall~t\,.
\ee
Differentiating this expression with respect to $t$ implies that $\p_t \hat P$ is proportional to a total divergence, in particular (as it happens in fluid mechanics, statistical mechanics or electrodynamics) to the divergence of its flux $\dot X\,\hat P$. Thus, the conservation law is $\p_t \hat P+ c \N_X\cdot(\dot X\,\hat P) =0$, for some constant $c$. From Eq.\ \Eq{gle0}, we can rewrite this expression as
\be\label{use1}
\p_t \hat P+ c \N_X\cdot(\eta\,\hat P) =0\,.
\ee
Integrating in time, one has
\ba
\hat P(X,x',t) &=& P_0(X,x')\nonumber\\
&&-c\int_{0}^t\rmd t'\, \N_X\cdot[\eta(t')\,\hat P(X,x',t')]\,,\label{use2}
\ea
where $P_0(X,x')=P(X,x',0)$ is the initial condition. For times prior to $t$, the functional $\hat P$ depends implicitly on the noise $\eta$. Plugging Eq.\ \Eq{use2} back into \Eq{use1}, we find
\ba
\p_t \hat P &=&-c \N_X\cdot(\eta\, P_0)+c^2 \N_X\cdot\left[\eta\int_{0}^t\rmd t' \N_X\cdot(\eta\hat P)\right]\nonumber\\
&=&0\,.\label{use3}
\ea
We take the average over the noise and call $P$ the stochastic average of the PDF [which depends on $\eta$, via Eq.\ \Eq{use1}]:
\be
P(x,x',t):=\langle \hat P\rangle_\eta\,.
\ee
The Gaussian statistics of the white noise now leads to the desired result. In fact, Eq.\ \Eq{wn} implies that the average of the first term in the right-hand side of \Eq{use3} vanishes ($X$ and $\eta$ are independent variables), while the bracket in the second term of \Eq{use3} is made of two contributions:
\ba
&&\int_{0}^t\rmd t' \N_X\cdot[\langle \eta(t)\eta(t')\rangle_\eta P(x,x',t')]\nonumber\\
&&\qquad+\int_{0}^t\rmd t'' [\eta(t'')\langle\eta(t)\N_X\cdot\hat P(X,x',t'')\rangle_\eta]\,,\nonumber
\ea
since $\hat P$ does include noise factors at time $t'$. The first correlation function gives $\de(t-t')$, and the second $\de(t-t'')$ for $t''<t'$. But $t>t''$ strictly, so the second contribution vanishes and, setting $c=1$ without loss of generality, we are left with
\be\label{BMde}
(\p_t-\kappa_1\N^2_x)\,P(x,x',t)=0\,,
\ee
where we have thrown away higher-order noise terms so that the Laplacian acts on the spatial coordinates $x$.

The diffusion coefficient $\kappa_1$ is measured in units of (length)$^2$/(time). Often it is effectively absorbed in the definition of a length variable
\be\label{BMlen}
\ell(t):=\sqrt{\kappa_1 t}\,.
\ee
In general, the diffusion coefficient is determined by the stochastic process underlying the diffusion equation (e.g., \cite{KTH,MeK00,Zas02}). Having averaged over noise, the average $\langle\cdot\rangle$ with respect to $\hat P$ and the one with respect to $P$ coincide, $\langle f(x,t)\rangle:=\int \rmd^D x P(x,x',t) f(x,t)$, and they shall be denoted in the same way without subscript. In particular, one can show that the mean-squared displacement (or second moment, or variance) from $x'=0$ is, in $D$ dimensions,
\be\label{msd0}
\langle X^2_{\textsc{bm}}(t)\rangle= 2D\kappa_1 t\,.
\ee

The Laplacian $\N^2$ in Eq.\ \Eq{BMde} acts on the $x$ dependence of the solution $P$, but in this particular case the latter depends only on the distance $r=|x-x'|$ between the initial and final points. In ordinary Euclidean space, the solution is the Gaussian distribution
\be\label{Gau}
P(x,x',t)=u_1(x,x',t):=\frac{\rme^{-\frac{|x-x'|^2}{4\ell^2(t)}}}{[4\pi\ell^2(t)]^{\frac{D}2}}\,,
\ee
where $|x-x'|^2=(x_1-x_1')^2+\cdots+(x_D-x_D')^2$ and the squared length $\ell^2$ is the dispersion of the Gaussian. Consistently with \Eq{hatu1}, the solution is a probability density with normalization
\be
\int_{-\infty}^{+\infty} \rmd^D x\, u_1(x,x',t)=1\,,
\ee
meaning that at all times or scales the pointwise test particle [represented by the delta initial condition $u_1(x,x',0)=\de(x-x')$] can be almost surely found somewhere.


\subsection{Spectral and walk dimension}\label{stsd}

So far we have interpreted the variable $t$, which has dimension $[t]=-1$, as a time parameter. Moving away from transport theory, in the context of spacetime theories $t\to\s$ is simply a length, representing the characteristic length scale $\ell$ [Eq.\ \Eq{BMlen} with $t\to\s$] at which one is probing the geometry. The Laplacian is then replaced by the curved Laplace--Beltrami operator in Wick-rotated spacetime.

When a stochastic process is staged on a smooth manifold with boundary, the return probability (or heat kernel, or partition function) $Z(\s)$ is simply defined as the functional trace of the solution $P$, i.e., its integral over the volume of the set when $x=x'$. Then, the first term in the Seeley--DeWitt expansion (e.g., \cite{Avr00,Kir01,Vas03}) is of the form $Z(\s)=\cV/(4\pi \kappa_1\s)^{-\ds/2}+\cdots$, where $\cV$ is the volume of the set. In the present case, however, this volume is infinite and it is customary to define the return probability as the trace per unit volume:
\be\label{BMrp}
\cP(\s)=\frac{Z(\s)}{\cV}=\frac{\int\rmd^Dx\,P(x,x,\s)}{\int\rmd^Dx}\,.
\ee
This definition coincides with the spatial average of the return probability density $P(x,x,\s)$. Since $P(x,x,\s)=u_1(x,x,\s)$ is constant in $x$ for a Brownian motion, the denominator in Eq.\ \Eq{BMrp} exactly cancels the divergence in the numerator and one ends up with $\cP(\s)=[4\pi\ell^2(\s)]^{-D/2}\propto \s^{-D/2}$. The spectral dimension is simply $\ds=D$ and it determines the decaying law of the return probability. In general, it is defined as
\be\label{dsdef}
\ds(\s) = -2\frac{\rmd\ln\cP(\s)}{\rmd\ln\s}\,,
\ee
or, more formally (due to a hidden divergence),
\be\label{dsdef2}
\ds(\s) := -2\frac{\rmd\ln Z(\s)}{\rmd\ln\s}\,.
\ee
Equation \Eq{dsdef2} is equivalent to the alternative definition $\tilde\ds := -\rmd\ln Z(\ell)/\rmd\ln\ell$ only for normal diffusion, but in general the correct one is \Eq{dsdef2} \cite{AO,RT,Wat85,HBA,KiL,bH}. Dimensionally, the mean-squared displacement $\langle X^2(\s)\rangle$ is always proportional to some squared length $\ell^2(\s)$. 

These definitions are meaningful in a spacetime geometry context provided $\s$ (or $\ell$) is interpreted as a measured length scale. Then, $\ds$ is the scaling law of the return probability when a test particle is left diffuse on the manifold. In this respect, one should notice that curvature and topology effects do modify the value $\ds=D$ at scales $\s$ larger than the characteristic curvature radius. The classic example is the sphere: at $\s\sim 0$, the probe locally feels a Euclidean plane, but as time passes the compact topology ``helps'' the particle going back to the initial point, so that $\ds\sim 0$ in the limit $\s\to+\infty$. Therefore, when the spectral dimension of spacetime is computed, the tacit understanding is that one is looking for the local geometric properties of the manifold, so that in all the cases where $Z$ is not expected to be a simple power law due to topology or curvature, it is more convenient to define
\be\label{dsdef3}
\ds = -2\lim_{\s\to 0^+}\frac{\ln Z(\s)}{\ln\s}\,.
\ee
Clearly, when $Z$ is a power law this expression coincides with \Eq{dsdef2} and $\ds$ is constant.

In more exotic scenarios of quantum gravity the effective diffusion equation can be very complicated \cite{frc4}, but at least in semiclassical regimes it must admit a positive semidefinite solution $P$. Otherwise, the operational definition of the spectral dimension \Eq{dsdef2} as the scaling of the return probability associated with a random walk would be lost. The requirement $P\geq 0$ was advocated, in particular, in \cite{frc4,CES}. 

The variance defines the so-called \emph{walk dimension}:
\be\label{defdw}
\langle X^2\rangle\propto \s^{{2}/{\dw}}\,.
\ee
From Eq.\ \Eq{msd0}, it follows that $\dw=2$ (normal diffusion) for ordinary Minkowski spacetime. In general, the spectral dimension $\ds$ precisely determines the anomalous scaling in $\s$, since there is a relation among Hausdorff, spectral, and walk dimensions \cite{bH}:
\be\label{dw}
\dw=2\frac{\dh}{\ds}\,.
\ee
Consistently, for Minkowski spacetime $\dh=D=\ds$. This relation holds for all fractals and can be understood from \Eq{defdw} in various ways \cite{bH}. A particularly simple one is to consider two random-walk systems characterized, respectively, by a length $\ell$ (the root-mean-square displacement $\ell\sim \sqrt{\langle X^2\rangle}$ of the walker) and a rescaled one $\ell\to\la \ell$. The energy density of states $\rho_\ell(E)$ is an extensive quantity proportional to the Hausdorff volume of the system, $\rho_\ell(E)\sim \ell^{\dh}$. 
 Therefore, under rescaling $\ell\to\la \ell$ one has
\be\label{E1}
\rho_{\la \ell}(E)=\la^{\dh} \rho_\ell(E)\,. 
\ee
On the other hand, from the Schr\"odinger equation one sees that energy $E$ is dimensionally conjugate to time, but from Eq.\ \Eq{defdw} ($\s\sim \ell^{\dw}$) there follows that 
\be\label{E2}
E_{\la \ell}=\la^{-\dw} E_\ell\,.
\ee
For probability to be conserved under rescaling, the densities of states in the original and rescaled systems must be related to each other by $\rho_{\la \ell}(E_{\la \ell})\rmd E_{\la E}=\rho_{\ell}(E_{\ell})\rmd E_{\ell}$, which implies, from Eq.\ \Eq{E2},
\be\label{E3}
\rho_{\la \ell}(E_{\la \ell}) =\la^{\dw}\rho_{\ell}(E_{\ell})\,.
\ee
The solution of Eqs.\ \Eq{E1}--\Eq{E3} is (removing the subscript $\ell$ from now on)
\be\label{rhoed}
\rho(E)\sim E^{{\dh}/{\dw}-1}=:E^{{\ds}/{2}-1}\,,
\ee
where in the last step we defined the spectral dimension in analogy with the ordinary Euclidean-space expression $\rho(E)\sim E^{D/2-1}$. Via quantum mechanics, one can then show (not so straightforwardly) that the definition in \Eq{rhoed} of $\ds$ coincides with the one from the diffusion equation.


\section{Multiscale spacetimes}\label{mfs}

A generic spacetime $\cM$ with scale-dependent geometry \cite{fra1,fra4} can be defined by equipping ordinary Minkowski spacetime with a measure $\rmd\vr(x)$, replacing the ordinary Lebesgue measure $\rmd^Dx$. A nontrivial metric structure independent of the measure can be added to describe manifolds with curvature, but we shall ignore it here, as we are interested in how a change of the differential structure (and of momentum space) affects the properties of spacetime. For technical reasons \cite{frc5,frc6}, it is convenient to concentrate on \emph{factorizable} measures,
\ba
\rmd \vr(x)&:=&\rmd^Dx\,v(x)=\rmd t\, v_0(t)\,\rmd{\bf x}\, v({\bf x})\nonumber\\
&:=&\rmd t\, v_0(t)\,\prod_{i=1}^{D-1}\rmd x^i\,v_i(x^i)\,,\label{genf}
\ea
where the $D$ functions $v_\mu$ can be all different. By definition, coordinates have dimension of lengths ($[x]=-1$ in momentum units). A further assumption is that the measure weight be positive semidefinite, $v_\mu=v_\mu(x^\mu)\geq 0$ for all $\mu=0,1,\dots,D-1$. An example of factorizable measure is the fractional one,
\bs\label{amea}\ba
\rmd\vr_\a(x)&=&\rmd^Dx\,v_\a(x)\,,\\
v_\a(x)&=&\prod_\mu v_\a(x^\mu):=\prod_\mu \frac{|x^\mu|^{\a_\mu-1}}{\Gamma(\a_\mu)}\,,
\ea\es
where $\G$ is the gamma function and $\a_\mu$ are $D$ real parameters (``fractional charges'') in the range $0<\a_\mu\leq 1$. In the simplest ``isotropic'' case, $\a_\mu=\a$ for all $\mu$. In general, we will call $\a$ the average fractional charge
\be\label{ava}
\a:=\frac{1}{D}\sum_\mu\a_\mu\,.
\ee

Fractional measures are of special interest in fractal geometry since they approximate the measure of random fractals \cite{frc1,RYS,RLWQ}. They are not unique (the position of the singularity parametrizes an infinite class, there exist nonfactorizable versions as in the Riesz integral, and so on \cite{frc1}) but they epitomize the simplest class of continuous measures with anomalous scaling law $\vr_\a(\la x)=\la^{\dh}\vr_\a(x)$, where $\dh$ is the Hausdorff dimension of spacetime. For Eq.\ \Eq{amea}, for instance,
\be\label{had}
\dh=\sum_\mu\a_\mu=D\a\leq D\,.
\ee
To obtain a scale-dependent geometry, it is sufficient to sum over a finite set of charges $\a$ \cite{fra4,frc2}. This step is important as soon as we want to employ these measures to describe realistic physical situations, where the dimension of spacetime is $D$ only at sufficiently large distances and low energies \cite{frc2}. Keeping factorizability, the multiscale measure weight is \cite{frc6}
\be\label{muf2}
v_*(x):=\prod_\mu v_*(x^\mu):=\prod_\mu\left[\sum_n g_n v_{\a_n}(x^\mu)\right]\,,
\ee
where $g_n\geq 0$ are dimensionful coefficients depending on a hierarchy of length scales $\ell_n>0$. For example, an isotropic spacetime with Hausdorff dimension $\dh\sim D$ in the infrared (IR) and $\dh\sim D\a_*<D$ in the ultraviolet (UV) is characterized by two charges $0<\a_1=\a_*<1$ and $\a_2=1$ and one fundamental length $\ell_1=\ell_*$ defining large and small scales. The measure weight then reads, for each coordinate,
\be\label{mufbin}
v_*(x)=1+\ell_*^{1-\a_*} v_{\a_*}(x)\,.
\ee
The volume of a $D$-ball of radius $R$ can be easily calculated and is of the form $\cV^{(D)}\sim \ell_*^D [\Omega_{D,1}(R/\ell_*)^D+\Omega_{D,\a_*} (R/\ell_*)^{D\a_*}]$, where $\Omega_{D,\a}$ is the fractional volume of a unit ball ($\Omega_{D,\a}=\Omega_{D,1}/[\Gamma(\a_*+1)]^D$ if centered at the origin) and we have thrown away off-diagonal terms in the integration. When $R\gg \ell_*$, the first term dominates and the ball volume scales as usual; otherwise, the second term dominates and the UV scaling is anomalous. In particular, if $\a_*=2/D$ spacetime has $\dh=2$ in the UV.

A more fundamental version of fractional spacetimes is endowed with complex fractional measures \cite{fra4,frc2,ACOS}. By a suitable choice of coefficients, one can construct measures with log oscillations, of which the present real-order measures are nothing but the zero mode. Log-oscillating measures describe spacetimes with discrete symmetries and a hierarchy of ultramicroscopic characteristic scales, smaller than those appearing in the real-order multifractional measure \Eq{muf2}. At the bottom of this tower, these is a fundamental length possibly identified with Planck's length \cite{ACOS}. The interest in these models lies in the phenomenological applications of such a scale hierarchy in quantum gravity, as well as on the fact that complex fractional measures better represent deterministic fractals \cite{NLM}. When defining the spectral dimension for these geometries, it is necessary to average over the log period of the measure in order to get a meaningful observable \cite{frc2}. The only surviving contribution is the zero mode, i.e., a power law with real fractional charge. Therefore, all information about the spectral dimension is encoded in the models considered in the present work, and we do not need to include log oscillations in the discussion.


\subsection{Laplacians}\label{lapl}

We consider three classes of multifractional spacetimes, each characterized by a different symmetry for the Lagrangian density and, hence, by a different Laplace--Beltrami (Laplacian in short) operator. The choice of symmetries in the action gives rise to physically inequivalent models with distinct predictions. In this paper, we concentrate exclusively on differences in their geometry by computing the associated stochastic processes and the spectral dimension. This is one first step in separating and characterizing the various versions of multiscale spacetimes. It is not exhaustive, and the next natural question is about physical predictions and whether some of these models are better motivated than the others. The answer lies beyond the scope of the present paper and will be the subject of future works.

In the remainder of this section, we leave the form of the weight $v(x)$ arbitrary while assuming the factorizability and positivity conditions.

\subsubsection{Weighted Poincar\'e symmetries}

With respect to the natural scalar product on $\cM$, one can construct a class of self-adjoint operators $\cK_{v,\g}$ of order $2\g$ generalizing the d'Alembertian \cite{frc4}. In particular, the second-order operator belonging to this class is \cite{frc3}
\be\label{ka}
\cK_v:=\eta^{\mu\nu}\cD_\mu\cD_\nu\,,\quad \cD_\mu:=\frac{1}{\sqrt{v(x)}}\,\p_\mu\left[\sqrt{v(x)}\,\,\cdot\,\right],
\ee
where $\eta_{\mu\nu}={\rm diag}(-,+,\cdots,+)$ is the Minkowski metric. A field action endowed with this operator is invariant under deformed Poincar\'e symmetries, where the generator algebra is the usual one in the free case, but the algebra elements do not generate the standard Poincar\'e transformations \cite{frc6}. In the interacting case, the algebra itself is deformed.

The eigenfunctions of $\cK_v$ are (here $k^2:=k_\mu k^\mu=-k_0^2+\sum_{i=1}^{D-1}k_i^2$)
\bs\label{E}\ba
\bE(k,x) &=& \frac{1}{\sqrt{w(k)v(x)}}\,\frac{\rme^{\rmi k\cdot x}}{(2\pi)^{\frac{D}{2}}}\,,\\
\cK_v\bE(k,x)&=&-k^2\bE(k,x)\,,
\ea\es
where $w$ is the measure weight of momentum space. The normalization is chosen so that one can write an invertible momentum transform:
\ba
\tilde f(k) &:=& \int_{-\infty}^{+\infty}\rmd^Dx\, v(x)\,f(x)\,\bE^*(k,x)\,,\label{fo1}\\
f(x) &=& \int_{-\infty}^{+\infty}\rmd^Dk\, w(k)\,\tilde f(k)\,\bE(k,x)\,.\label{fo2}
\ea
The multiscale generalization of the Dirac distribution is then
\ba
\de_v(x,x') &:=&\frac{\de(x-x')}{\sqrt{v(x)v(x')}}\nonumber\\
&=& \int\rmd^Dk\, w(k)\,\bE^*(k,x)\bE(k,x')\,,\label{dev}
\ea
with a similar expression for $\de_w(k,k')$. The operator \Eq{ka} was considered in \cite{frc1,frc2,frc3,fra6,frc4,frc5,fra7,frc6}.

\subsubsection{Ordinary Poincar\'e symmetries}

In another scenario \cite{fra1,fra2,fra3,frc1,frc2} the Lagrangian density possesses ordinary Poincar\'e symmetries and the Laplace--Beltrami operator is simply 
\be\label{ka1}
\B:=\eta^{\mu\nu}\p_\mu\p_\nu\,,
\ee
which is not self-adjoint with respect to the scalar product with measure weight $v(x)$. In fact,
\be\label{boda}
\B^\dagger=\check{\cK}_v:=\frac{1}{v(x)}\,\B\left[v(x)\,\cdot\,\right]\,.
\ee
Hence, a (for instance) scalar field model with quadratic-like kinetic term $-\p_\mu\phi\p^\mu\phi/2$ is inequivalent to the one with Gaussian-like term $\phi\B\phi/2$. In this case, the field theory has some ordering prescription and there is no direct definition of a self-adjoint momentum operator (and there may appear complications with the microcausality structure of the theory \cite{frc2}). Correspondingly, it is not clear whether an invertible momentum transform exists. Still, the operators \Eq{ka1} and \Eq{boda} will be of interest in what follows.

\subsubsection{\texorpdfstring{$q$}{}-Poincar\'e symmetries}

The action measure can be recast as $\rmd\vr=\rmd^Dq$, where $\rmd q^\mu=\rmd x^\mu v_\mu(x^\mu)$. Notice that this writing, as well as the most general one $q(x)=\vr(x)$, is equivalent to Eq.\ \Eq{amea} only in the sense of distributions. In the case of a fractional spacetime with fixed dimensionality, for each direction one has
\be
q^\mu(x^\mu)=\vr_\a(x^\mu)=\frac{{\rm sgn}(x^\mu)|x^\mu|^{\a_\mu}}{\Gamma(\a_\mu+1)}\,.
\ee

Obviously, the Hausdorff dimension is not affected by a change of variables \cite{frc1}, and it is Eq.\ \Eq{had}, $\dh=D\a$. This happens because \emph{by definition} the momentum space of the theory is conjugate to position space in $x$, not in $q$. Consequently, the units of the $x$ coordinates are $[x^\mu]=-1$ while the $q$'s have anomalous scaling, $[q^\mu]=-\a_\mu$. This feature guarantees that the theory is not trivial \cite{fra7} even if one defines it such that in $q$ variables it is formally identical to the ordinary one, including the Laplace--Beltrami operator
\ba
\B_{q(x)}=\eta^{\mu\nu}\frac{\p}{\p q^\mu(x)}\frac{\p}{\p q^\nu(x)}
=\eta^{\mu\nu}\frac{1}{v_\mu(x)}\p_\mu\left[\frac{1}{v_\nu(x)}\p_\nu\,\cdot\,\right],\nonumber\\\label{boq}
\ea
which is, therefore, self-adjoint under suitable boundary conditions. In the last equation we omitted the indices in the arguments of $q$ and $v$. 

The measure, the Lagrangian, and the action as a whole possess ``$q$-Poincar\'e'' symmetries; i.e., they are Lorentz and translation invariant under transformations over the $q$'s, which are then nonlinear transformations over the $x$'s \cite{frc1,frc2}:
\be
{q'}^\mu(x')=\Lambda_{\ \nu}^\mu q^\nu(x)+a^\mu\,.
\ee 


\subsection{Previous ansatz for the diffusion equation}\label{old}

The explicit calculation of the spectral dimension $\ds$ was performed in \cite{frc1} for fixed dimensionality and in \cite{frc2,frc4,frc6} for multifractional spacetimes, in the case of the weighted Laplacian \Eq{ka}.\footnote{A heuristic estimate of $\ds$ for the ordinary Laplacian can be found in \cite{fra2}.} We recall here the result for an integer-order diffusion equation and an isotropic fractional measure ($\a_\mu=\a$, fixed dimensionality). The multiscale anisotropic case is more complicated and adds nothing to the main point we wish to make here.

The ansatz adopted for the diffusion equation was based on two assumptions: (a) since $\s$ is a fictitious variable, the diffusion operator $\p_\s$ should not reflect the differential structure of fractional spacetime and can be left unchanged; (b) as the spatial generator in the ordinary diffusion equation \Eq{BMde} coincides with the self-adjoint Laplacian appearing in the action, so should it be in the fractional case. This singles out the diffusion equation
\be\label{fde0}
(\p_\s-\kappa_1\cK_v) \tilde P(x,x',\s)=0\,,\quad \tilde P(x,x',0)=\de_v(x,x')\,.
\ee
The first problem arises because the solution $\tilde P(x,x',\s)=u_1(x,x',\s)/\sqrt{v(x)v(x')}$ is not normalized to 1. It is easy to see that $P:= \sqrt{v(x')/v(x)}\tilde P=u_1/v(x)$ obeys the same initial condition, is normalized to 1, and is a solution of 
\be\label{fde1}
(\p_\s-\kappa_1\check{\cK}_v)P(x,x',\s)=0\,,
\ee
rather than of Eq.\ \Eq{fde0} \cite{frc4}. The operator $\check{\cK}_v$ is not self-adjoint, which requires us to abandon assumption (b). This is not an issue, since Fokker--Planck equations are in general not self-adjoint \cite{Zwa01}. One should now explain, however, why the spatial generator carries weights $v$ instead of $\sqrt{v}$, but for the time being we just ignore this point and move on.

In analogy with Eq.\ \Eq{BMrp}, we maintain the definition of return probability as the functional trace of the solution $P$ per unit volume:
\bs\label{frp1}\ba
\cP(\ell)&=&\frac{Z(\ell)}{\cV_{\rm H}}=\frac{1}{\cV_{\rm H}}\int\rmd\vr(x)\,P(x,x,\ell)\,,\\
\cV_{\rm H} &:=&\int\rmd\vr(x)=\int \rmd^D x\, v(x)\,.
\ea\es
In fractional spacetimes, however, the spacetime integral in the denominator does not cancel the one in the numerator. In fact, the return probability reads
\be\label{bad}
\cP(\ell)=\frac{\int\rmd^Dx}{\int\rmd^Dx\, v(x)}\frac{1}{(4\pi\ell^2)^{\frac{D}2}}\,.
\ee
The ratio of the two spatial integrals is a divergent, dimensionful constant. To get rid of it, one must first extract all the dimensional dependence. This is done by defining the dimensionless coordinates $\tilde x^\mu = x^\mu/\ell$, so that
$\cP(\ell) =C \ell^{-D\a}$. The constant $C$ is formally divergent but it can be regularized \cite{frc4} and, since it is dimensionless, be thrown away. It is completely immaterial in the definition \Eq{dsdef}, which would give
\be\label{dsda}
\ds=D\a\,.
\ee

Unfortunately, there is an issue with this procedure, which we can state in several ways. One is to notice that the regularization trick would introduce a degree of arbitrariness in the return probability which, if the underlying stochastic process is well defined, should not be there. In other words, the definitions \Eq{dsdef} and \Eq{dsdef2} are inequivalent in this case, while they should always agree because the overall volume prefactor in $\cP$ does not depend on diffusion time. This would suggest that we use \Eq{dsdef2} or, equivalently, that we define $\cP$ as $\cP=Z/(\int\rmd^Dx)$ instead of Eq.\ \Eq{frp1}, thus modifying the result of the spectral dimension to $\ds=D$. This is not a fractal, since $\ds=D\geq \dh$ (for fractals, $\ds\leq \dh$). The interpretation of $\cP$ as the spatial average of $P$ would also be lost. On the other hand, one might wave away worries that the overall constant in the return probability \Eq{bad} is regularization dependent on account of two observations. First, in scenarios with fixed dimensionality the regularization only affects the normalization of the PDF, not its signature. Second, when $\ds$ is multiscale, transient regimes between asymptotic plateaux may also depend on the regularization, which is a known feature of multiscale systems including field theories of quantum gravity \cite{frc4}.

Another way to see the problem is to recall that the heat kernel for fractals is of the form
\be\label{Zfr}
Z(\ell)=\frac{\cV_{\rm H}}{(4\pi\ell^2)^{\ds/2}}+\cdots\,,
\ee
where $\cV_{\rm H}=L^{\dh}$ is the spectral volume of the set, $L$ is the characteristic spectral length determined by the Laplacian (roughly, by the periodicity of its eigenfunctions) and $\dh$ is the Hausdorff dimension of the set \cite{Akk12,Akk2}. On the other hand, the divergent contribution from the heat kernel $Z$ in \Eq{bad} is the ordinary integer volume, not the Hausdorff volume. This version of the problem can be softened by recalling that the first Seeley--DeWitt coefficient in the heat kernel is quite generically equal to the spectral volume but, to the best of our knowledge, it has not been proven as a strict rule for nonsmooth sets. It is not obvious, however, why this should not be the case also for fractional spacetimes, which are a hybrid between continuum geometries and genuine fractals.

A third manifestation of the problem becomes apparent when one identifies its origin in the normalization of plane waves \Eq{E}: the phases $\bE(k,x)$ are normalized per unit integer volume, not per fractional volume. This is apparent from Eq.\ \Eq{dev}. In yet other words, the number of states is not the fractional volume in Eq.\ \Eq{frp1}
but, rather, $\de(0)/v(x)$. In fact, we can define the density of states of a system with Hamiltonian eigenstates $\psi_k(x)$ and energy $E_k$ as
\be\label{rhoE}
w(E)\,\rho(E):= \int\rmd^Dk\, w(k) |\psi_k(x)|^2\de(E-E_k)\,,
\ee
where $w(E)$ is the energy measure weight. Then, the total number of states is
\be\label{resu}
\int \rmd E\,w(E)\, \rho(E)= \int\rmd^Dk\, w(k) |\psi_k(x)|^2 = \frac{\de(0)}{v(x)}\,,
\ee
where in the last step we used the fact that the states $\psi_k$ always contain a factor of the form $1/\sqrt{v(x)w(k)}$ [as in the free particle case, where $\psi_k(x)=\bE(k,x)$]. In the bra-ket formalism, this is tantamount to starting with the resolutions of the identity
\bs\ba
\mathbbm{1}_k &=& \frac{1}{(2\pi)^{D/2}}\int \rmd^D k\, w(k) |k\rangle\langle k|\,,\\
\mathbbm{1}_x &=& \frac{1}{(2\pi)^{D/2}}\int \rmd^D x\, v(x) |x\rangle\langle x|\,,
\ea\es
so that $|x\rangle = (2\pi)^{-D/2}\int \rmd^D x' v(x') |x'\rangle\langle x'|x\rangle$ and one finds $\langle x'|x\rangle=(2\pi)^{D/2}\de_v(x,x')$. Since $(2\pi)^{D/2}\de_v(x,x')=\langle x'|\mathbbm{1}_k|x\rangle$, the right-hand side of \Eq{resu} can be recast as $\langle x|x\rangle=\de(0)/v(x)$, where $\de(0)$ is the volume of integer momentum space. It is the integer volume because the eigenfunctions $\psi_k(x)$ always cancel the weight factor $w(k)$. Therefore, each infinitesimal hypercube $\rmd^Dk$ (or $\rmd^Dx$) contains $w(k) \rmd^Dk/w(k)=\rmd^Dk$ localized states (respectively, $\rmd^Dx$), as in the usual case.


Ultimately, the problem lies in the diffusion equation itself, Eq.\ \Eq{fde1}, which was simply assumed. Starting from the Langevin equation we will get, instead, a different diffusion equation not only derived in closer conformity with the standard case of ordinary space, but also capable of overcoming the difficulties outlined above.


\section{Diffusion in fractional spacetimes}\label{diffs}


\subsection{Weighted Laplacian}\label{abm}

The natural extension of classical mechanics to multiscale spacetimes is obtained by replacing ordinary time derivatives with weighted derivatives $\cD_t$ \cite{frc5}. For instance, a free particle in a quadratic potential has action
\be\nonumber
S=\int\rmd t\, v_0(t)\left[\frac{1}{2}m\left(\cD_t x\right)^2 - \frac{1}{2} m {\om^2} x^2 \right]\,,
\ee
with equation of motion
\be\nonumber
m\cD^2_t x +m{\om^2} x=0\,.
\ee
Similarly, it is natural to define the Langevin equation for a random variable $X$ in fractional or multiscale space as
\be
m \cD_t^2 X+m\g \cD_t X+U'(X)=F\,.
\ee
Switching to the spacetime interpretation (A)--(C) spelled out in the introduction, one ends up with the stochastic equation
\be\label{lef}
m \cD_\s^2 X+m\g \cD_\s X+U'(X)=F\,,
\ee
where we now use diffusion time according to assumption (B). However, now assumption (C) is less transparent than in the standard case, since the derivative $\cD_\s$ contains a weight $v(\s)$ for an abstract evolution parameter and, \emph{a priori}, it is not obvious whether this weight should have the same functional form as the weight $v_0(t)$. For the time being, we leave the form of $v(\s)$ unspecified, calling $1-\b=[v(\s)]$ its scaling dimension. In particular, in the fractional case we set
\be\label{vs}
v(\s)=v_\b(\s):=\frac{\s^{\b-1}}{\Gamma(\b)}\,,\qquad \b>0\,,
\ee
with $\b$ not necessarily equal to the fractional charge $\a_0$.

We now make the field redefinition
\be\label{YX}
Y(\s):= \sqrt{v(\s)}\, X(\s)\,,\quad \xi(\s):=\frac{1}{m\g}\sqrt{v(\s)}\,F(\s)\,,
\ee
and Eq.\ \Eq{lef} reduces to \Eq{le00} in the presence of a nonautonomous potential $W=\sqrt{v(\s)}U$ and a noise $\xi$:
\be\label{lef2}
\frac{1}{\g}\p^2_\s Y+\p_\s Y+ \frac{W'(\s,Y)}{m\g}=\xi\,.
\ee
The resemblance with Eq.\ \Eq{le00} is only superficial, since measure units differ: the effective variable $Y$ has scaling dimension $[Y]= -(1+\b)/2$, while $[\xi]=(1-\b)/2$. Notice that if $X$ is a random variable, so is $Y$, since $v(\s)$ is a deterministic function. Also, at the initial time $\s=0$ one takes the expression $Y(0)=\lim_{\s\to 0}\sqrt{v(\s)}\, X(\s)$ formally finite, even if $v(0)$ may diverge.

A natural assumption is that $F$ is a Gaussian white-noise field but in a fractional sense, i.e., with correlation $\langle F(\s) F(\s')\rangle_F\propto \de_v(\s,\s')$, where the proportionality coefficient has dimension (mass)$^{4-\b}$ ($\langle FF\rangle$ has scaling dimension 4 and $[\de_v]=\b$). It is more instructive, however, to consider the general case
\bs\label{corr}\ba
&&\langle F(\s) F(\s')\rangle_F= (m\g)^2 \kappa_{\b,\nu}\, \s^{\nu-1} \de_v(\s,\s')\\
&&\quad \Rightarrow\quad \langle \xi(\s) \xi (\s')\rangle_\xi= \kappa_{\b,\nu}\, \s^{\nu-1} \de(\s-\s')
\ea\es
for all $\s$ and $\s'$, where $\nu$ is a real parameter and $\kappa_{\nu,\b}$ is a constant with dimension $[\kappa_{\b,\nu}]=\nu-\b-1$. Repeating the same calculation as in Sec.\ \ref{euc}, one finds an expression for the probability density $\hat u$ (we reserve the symbol $\hat P$ for later) satisfying an equation of the form \Eq{use3}, with $X$ replaced by $Y$. Taking the average over the noise and calling $u(y,y',\s):=\langle \hat u\rangle_\eta$, we end up with
\be\label{tddc0}
(\p_\s-\kappa_{\b,\nu}\s^{\nu-1}\N^2_{y})\,u(y,y',\s)=0\,,
\ee
where $y:=\sqrt{v(\s)} x$ actually contains a dependence on $\s$. Restoring the coordinates $x$, we finally obtain
\be\label{tddc}
[\p_\s-\kappa(\s)\N^2_x]\,u_{\b,\nu}(x,x',\s)=0\,, \quad \kappa(\s):=\kappa_{\b,\nu} \frac{\s^{\nu-1}}{v(\s)}\,.
\ee
In particular, for the fractional measure \Eq{vs} $\kappa(\s)=\kappa_{\b,\nu} \Gamma(\b)\s^{\nu-\b}$. Notice that the solution of the diffusion equation \Eq{tddc} is not the solution of \Eq{tddc0}, since the function $u[\sqrt{v(\s)} x, y',\s]$ is neither correctly normalized to 1 nor well defined for initial points at $|x'|<\infty$. The actual fractional solution $u_{\b,\nu}(x,x',\s)$ is a Gaussian PDF proportional to $u(y, y',\s)$:
\be\label{unb}
u_{\b,\nu}(x,x',\s)=\frac{\rme^{-\frac{|x-x'|^2}{4\ell^2(\s)}}}{[4\pi\ell^2(\s)]^{\frac{D}2}}\,,
\ee
where the dispersion $\ell^2$ is
\ba
\ell^2(\s) &:=& \bar{\ell}^2+\int^\s \rmd\s'\, \kappa(\s')= \bar{\ell}^2+\kappa_{\b,\nu} \int^\s \rmd\s'\,\frac{{\s'}^{\nu-1}}{v(\s')}\,,\nonumber\\\label{ell2}
\ea
and we allowed for an additive constant $\bar{\ell}^2$. The initial condition $u_{\b,\nu}(x,x',0)=\de(x-x')$ imposes that $\ell^2(0)=0$, which fixed $\bar{\ell}^2$ depending on the measure $v(\s)$. In the fractional case \Eq{vs}, one has that $\bar{\ell}=0$ and
\be\label{ell22}
\ell^2(\s)= \kappa_{\b,\nu} \frac{\Gamma(\b)}{1+\nu-\b}\s^{1+\nu-\b}\,.
\ee
If we had set $\bar{\ell}\neq 0$, we would have obtained a Gaussian initial spread $u_{\b,\nu}(x,x',0)=\exp[-|x-x'|^2/(4\bar{\ell}^2)]/(4\pi\bar{\ell}^2)^{D/2}$ of width $\bar{\ell}$, a situation mimicking the idea of a particle on a fuzzy manifold with a minimal length \cite{SmS1,SmS2,SmS3,SSM,Rin09,MoN,NiN}. We will come back to this point in Sec.\ \ref{mus}, where a nonvanishing $\bar{\ell}$ is necessary in certain multiscale scenarios.

Equation \Eq{tddc} is not yet in its final form, since we have not discussed the normalization of the solution in a multiscale spacetime. In spacetimes with nontrivial measure, the normalization of the PDF is done with respect to the total measure weight $v(x)$. The only way to get the result and maintain both the Langevin-equation interpretation and the delta initial condition is to define the fractional PDF as
\be\label{Puv}
P_{\b,\nu}(x,x',\s)=\frac{u_{\b,\nu}(x,x',\s)}{v(x)}\,,
\ee
which is the solution of
\ba
[\p_\s-\kappa(\s)\check{\cK}_v]P_{\b,\nu}(x,x',\s)&=&0\,,\nonumber\\
P_{\b,\nu}(x,x',0)&=&\de_v(x,x')\,.\label{tddcf}
\ea
Since $u_{\b,\nu}$ obeys the self-similarity relation
\ba
&&u_{\b,\nu}[\la^{(1+\nu-\b)/2}x,\la^{(1+\nu-\b)/2}x',\la\s]\nonumber\\
&&\qquad=\la^{-D(1+\nu-\b)/2}u_{\b,\nu}(x,x',\s)\,,\label{us}
\ea
$P_{\b,\nu}$ is self-similar with scaling law
\ba
&&P_{\b,\nu}[\la^{(1+\nu-\b)/2}x,\la^{(1+\nu-\b)/2}x',\la\s]\nonumber\\
&&\qquad=\la^{-D\a(1+\nu-\b)/2}P_{\b,\nu}(x,x',\s)\,.
\ea
Quantum mechanics motivates the appearance of the operator $\check{\cK}_v$ in the diffusion equation \Eq{tddcf} independently (see Appendix \ref{qmbil}). It also determines, independently from the Langevin-equation approach, the natural diffusion time \Eq{ell2} (see Appendix \ref{qmtime}).

\subsubsection{Factorizable-spacetime Brownian motions}\label{fsbms}

\begin{figure}
\centering
\includegraphics[width=8.4cm]{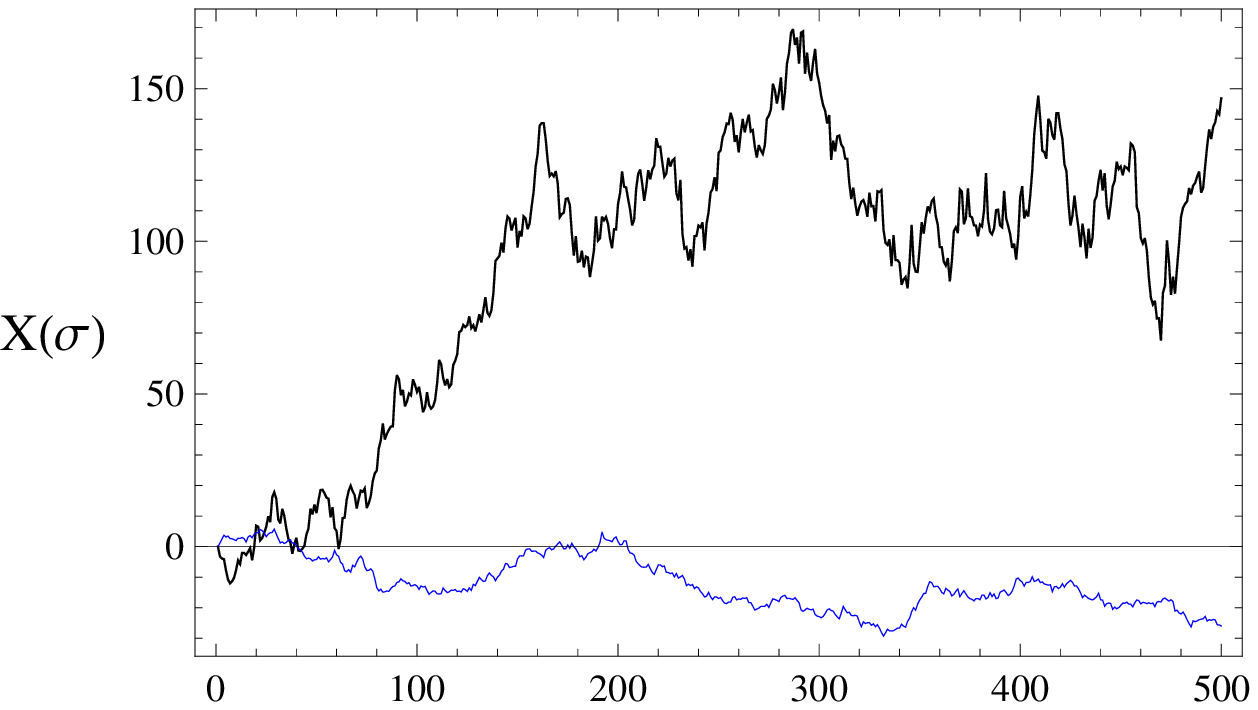}\\
\includegraphics[width=8.4cm]{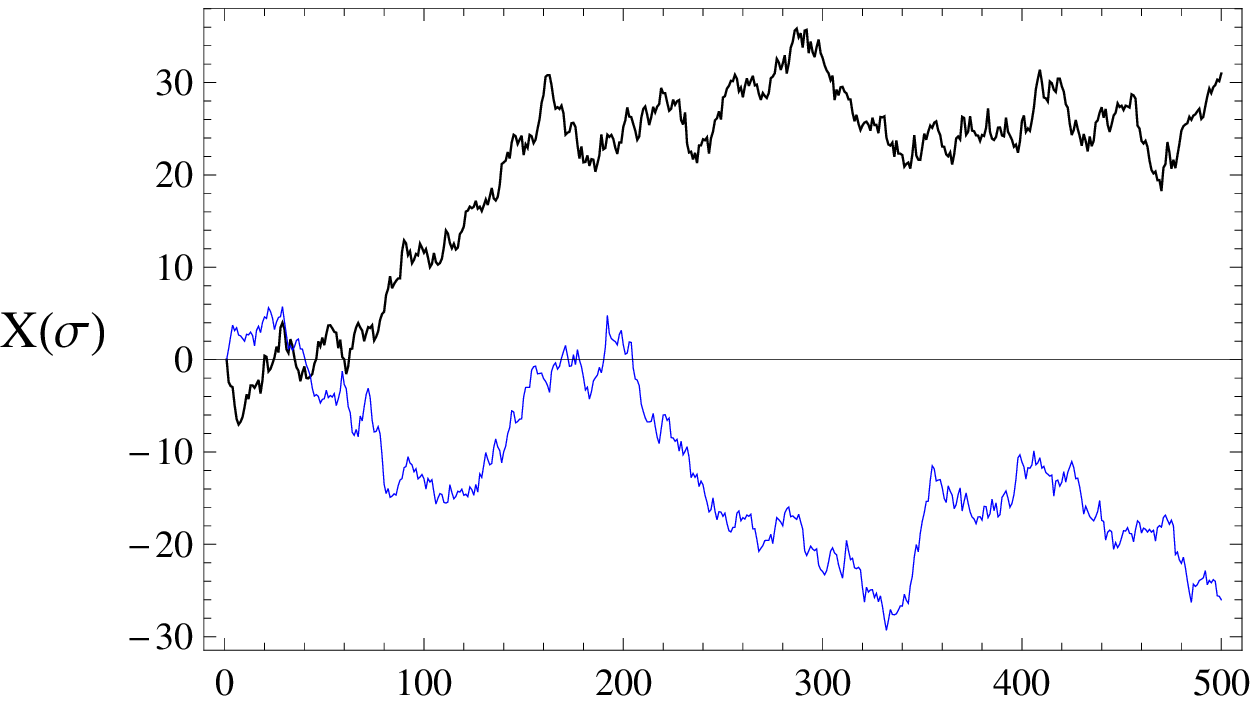}\\
\includegraphics[width=8.4cm]{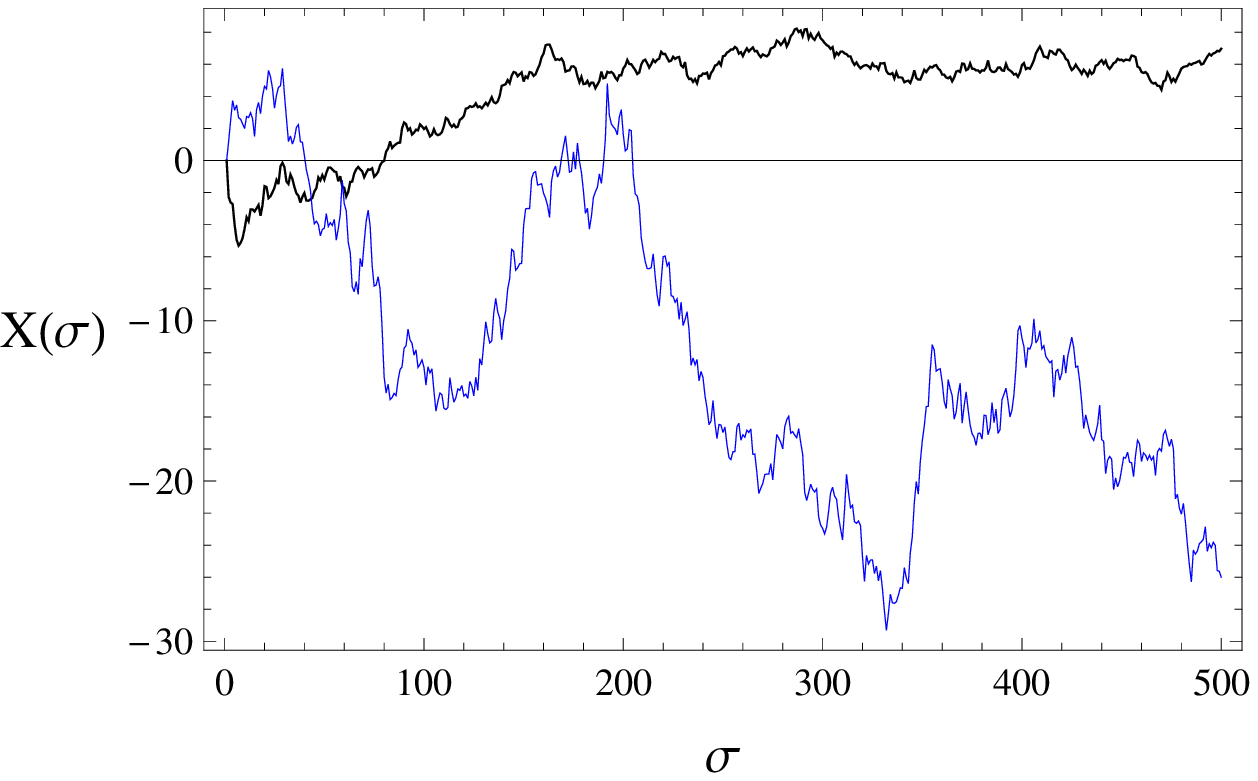}
\caption{\label{fig1} Dark (black) curve: Example of trajectory of the process \Eq{fsbm} in a fractional spacetime with weighted Laplacian, diffusion measure weight \Eq{vs}, $\nu=1$, and $\b=1/2$ (top panel), $\b=1$ (middle panel, ordinary Brownian motion), and $\b=3/2$ (bottom panel). Light (blue) curve: Example of trajectory of ordinary Brownian motion in ordinary spacetime, plotted for reference (notice the different scaling of the vertical axes).}
\end{figure}

Set first $\nu=1$. From Eqs.\ \Eq{tddc0} and \Eq{YX}, it follows that $X$ is the stochastic process
\be\label{fsbm}
\boxd{X_{\textsc{fsbm-}v}(\s)=\frac{X_{\textsc{bm}}(\s)}{\sqrt{v(\s)}}\,.}
\ee
This is a plain Brownian motion, but in a fractional or, in general, factorizable multiscale spacetime. We label the factorizable-spacetime Brownian motion (FSBM) in this model of multiscale geometry with weighted Laplacian as FSBM-$v$. Some trajectories for the weight \Eq{vs} are shown in Fig.\ \ref{fig1}. The raggedness of the curve and the drift from the average decrease when $\b$ increases. 

Equations \Eq{sca0}, \Eq{vs}, and \Eq{fsbm} imply the self-similarity property
\be\label{scaf}
X_{\textsc{fsbm-}v}(\la\s)=\la^{\frac{2-\b}2} X_{\textsc{fsbm-}v}(\s)\,.
\ee
The mean-squared displacement of FSBM is anomalous,
\be\label{msd}
\langle X^2_{\textsc{fsbm-}v}(\s)\rangle=\int\rmd^Dx\,v(x)\, P_{\b,\nu}(x,x',\s)\,x^2\propto \ell^2(\s)\,,
\ee 
and, for the fractional case $\ell^2(\s)\propto \s^{2-\b}$, one has subdiffusion when $0<\b<1$. Moreover, increments are uncorrelated but not stationary, since their distribution does not depend on the time interval only:
\be
\langle [X_{\textsc{fsbm-}v}(\s)-X_{\textsc{fsbm-}v}(\s')]^2\rangle\neq \langle X^2_{\textsc{fsbm-}v}(\s-\s')\rangle\,.
\ee

\begin{figure}
\centering
\includegraphics[width=8.4cm]{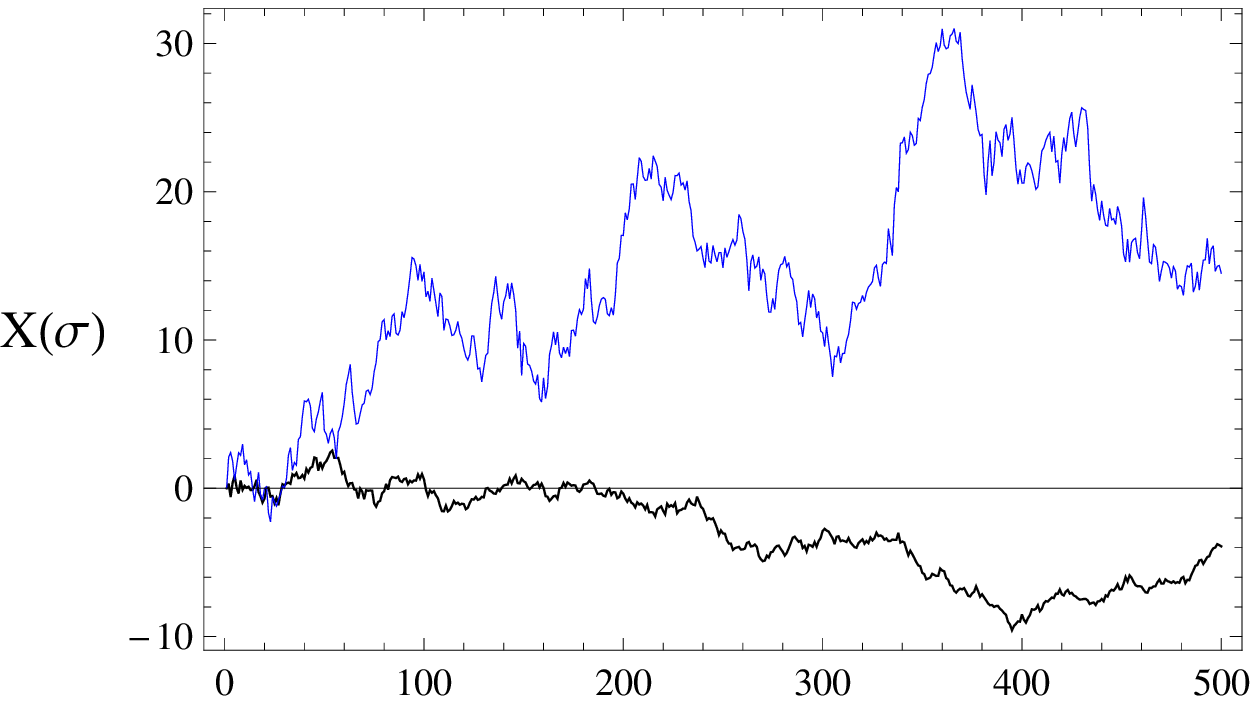}\\
\includegraphics[width=8.4cm]{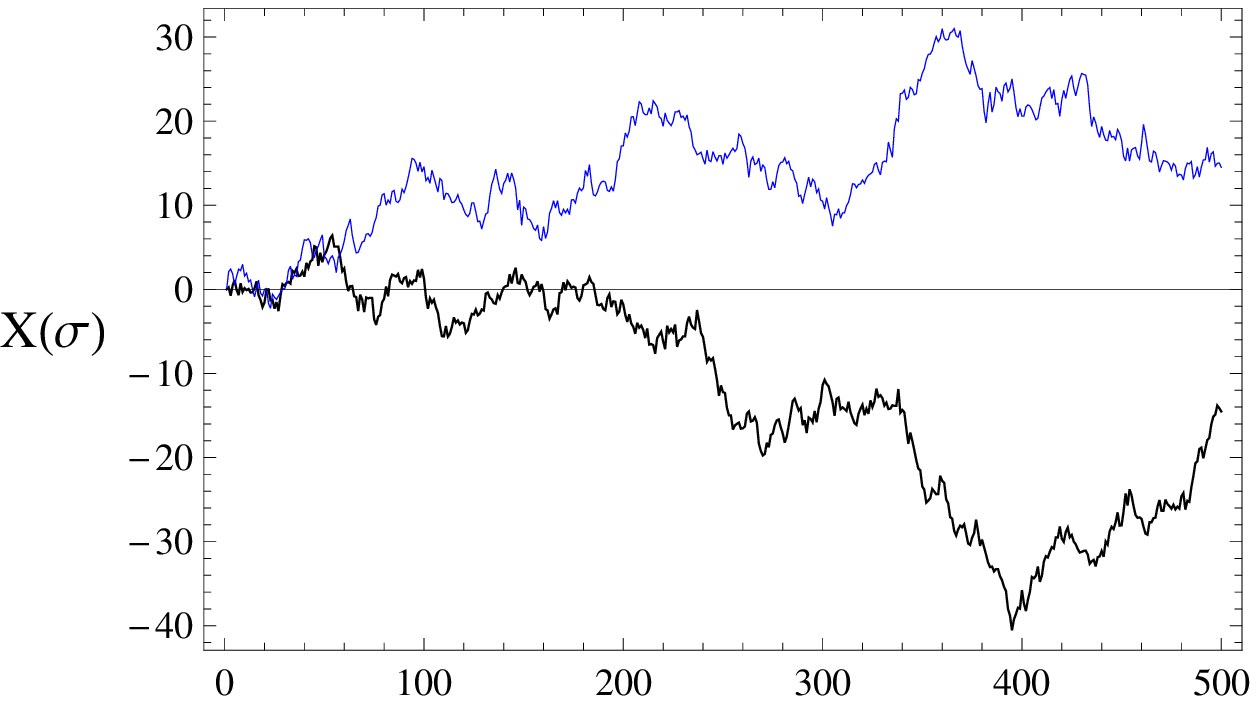}\\
\includegraphics[width=8.4cm]{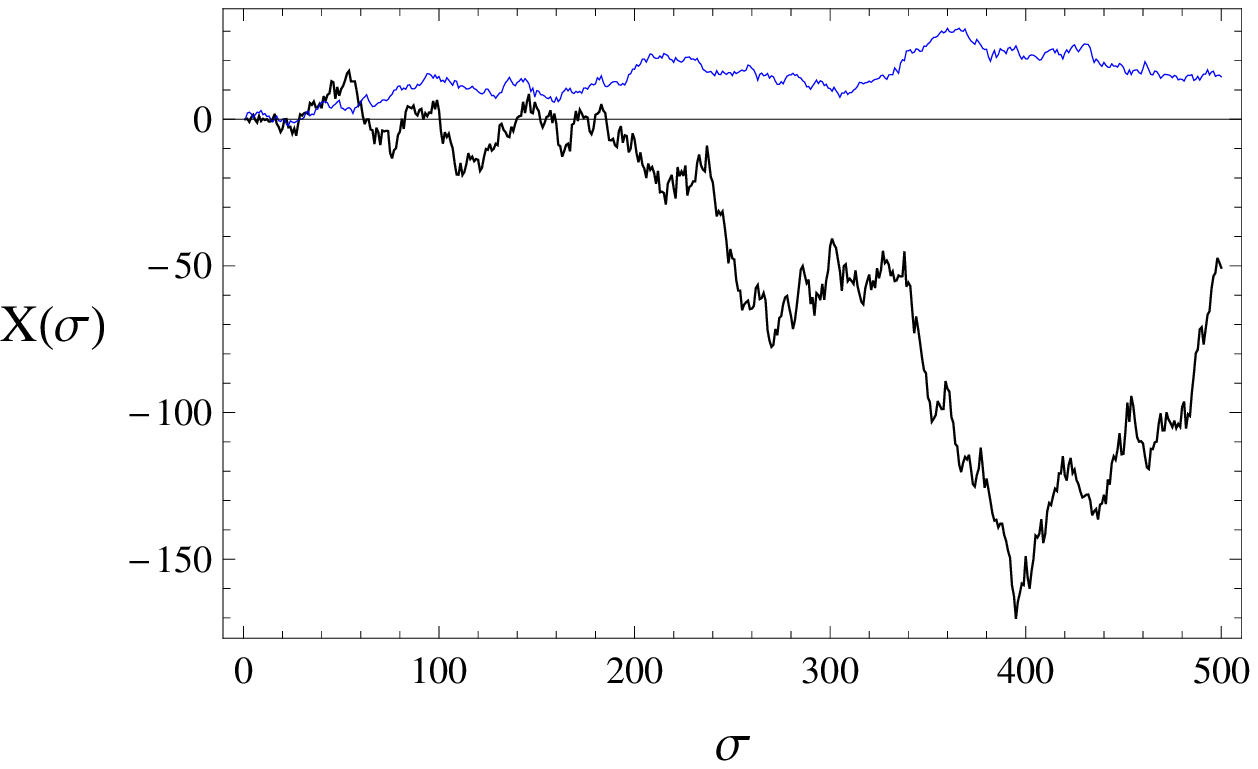}
\caption{\label{fig2} Dark (black) curve: Example of trajectory of scaled Brownian motion \Eq{scabm} in ordinary spacetime with $\nu=1/2$ (top panel), $\nu=1$ (middle panel, ordinary Brownian motion), and $\nu=3/2$ (bottom panel). The same trajectories represent the process \Eq{opsbm} in a fractional spacetime with ordinary Laplacian with $\nu=1$ and, respectively, $\b=3/2$, $1$, $1/2$. Light (blue) curve: Example of trajectory of ordinary Brownian motion in ordinary spacetime, plotted for reference (notice the different scaling of the vertical axes). The code implements the algorithm presented in Eq.\ (A2) of \cite{LiM02}.}
\end{figure}

If $\nu\neq 1$, set $\xi=\s^{(\nu-1)/2}\eta$, where $\eta$ is a Gaussian white noise \Eq{wn}. Then, the Langevin equation [from \Eq{lef2}] $\p_\s Y=\xi$ becomes
\be\label{gleb}
\p_\s X_{\textsc{sbm}}(\s) = \s^{\frac{\nu-1}2}\eta(\s)\,,
\ee 
where we called $Y(\s)=X_{\textsc{sbm}}(\s)$ the solution. This is nothing but a \emph{scaled Brownian motion} (SBM) \cite{LiM02,MeK04,Sok12}, defined as
\be\label{scabm}
X_{\textsc{sbm}}(\s) := X_{\textsc{bm}}(\s^\nu)\,.
\ee
Its self-similarity property is $X_{\textsc{sbm}}(\la\s)=\la^{\nu/2} X_{\textsc{sbm}}(\s)$.  Just like FSBM, SBM is Markovian, since the scale transformation $\s\to\s^\nu$ preserves time ordering for $\nu>0$ \cite{LiM02}. Again, increments are nonstationary, $\langle [X_{\textsc{sbm}}(\s)-X_{\textsc{sbm}}(\s')]^2\rangle=\langle [X_{\textsc{bm}}(\s^\nu)-X_{\textsc{bm}}({\s'}^\nu)]^2\rangle= \langle X^2_{\textsc{bm}}(\s^\nu-{\s'}^\nu)\rangle\neq \langle X^2_{\textsc{sbm}}(\s-\s')\rangle$. Figure \ref{fig2} shows the effect of the time rescaling: as $\nu$ decreases, the trajectory becomes smoother.

Equations \Eq{fsbm} and \Eq{scaf} are replaced by a fractional-spacetime scaled Brownian motion (FSSBM)
\be
\boxd{X(\s)=X_{\textsc{fssbm}}(\s):=\frac{X_{\textsc{sbm}}(\s)}{\sqrt{v(\s)}}\,,}
\ee
with scaling law
\be
X_{\textsc{fssbm}}(\la\s)=\la^{\frac{1+\nu-\b}2} X_{\textsc{fssbm}}(\s)\,.\label{fssbmv}
\ee
Its variance is proportional to $\ell^2$, which for the fractional case \Eq{vs} is
\be\label{msdn}
\langle X^2_{\textsc{fssbm}}(\s)\rangle \propto \s^{1+\nu-\b}\,.
\ee

We notice that, in an ordinary spacetime, Eq.\ \Eq{tddc} with time-dependent diffusion coefficient can describe two different stochastic processes. One is SBM. Another is fractional Brownian motion (FBM) \cite{BaA,MaV} (see also \cite{WaL,Wan92,WaT,Lut01,LiM02,MeK04,JeM,Sok12}). In this case, the GLE features a Weyl fractional integral of order $\g=(2+\nu-\b)/2$ and reads $\p_\s X_{\textsc{fbm}}(\s) = \p_\s[{}_{-\infty}I^\g\eta](\s)$. Scaled and fractional Brownian motion are \emph{twins} \cite{Sok12}; i.e., they share exactly the same diffusion equation \Eq{tddc} and the mean-squared displacement. However, they are physically quite different, since FBM is non-Markovian because its future evolution depends also on past states, due to its definition via a nonlocal operator. Also, FBM has stationary increments, while SBM has not. From the point of view of fractional spaces, there is no twin problem and no ambiguity. In fact, the natural Langevin equation is
\be\label{laef}
\p_\s[\sqrt{v(\s)}\,X]=\xi\,,
\ee
which is neither the one of SBM nor of FBM. Furthermore, we are dealing with a Markovian process, as the derivation of the diffusion equation has shown. The Markovian stochastic process associated with Eqs.\ \Eq{laef} and \Eq{tddc} is unambiguously identified with FSSBM, or with FSBM-$v$ if $\nu=1$.

\subsubsection{Spectral and walk dimension}\label{spewaf}

For a generic measure $v(\s)$, the spectral dimension is
\ba
\ds(\s) &\ \stackrel{\text{\tiny \Eq{dsdef2}}}{=}\ & -2\frac{\rmd\ln Z(\ell)}{\rmd\ln\ell^2} \frac{\rmd\ln\ell^2(\s)}{\rmd\ln\s}\nonumber\\
        &\ \stackrel{\text{\tiny \Eq{unb}}}{=}\ & D\frac{\rmd\ln\ell^2(\s)}{\rmd\ln\s}\label{newdsq}\\
        &\ \stackrel{\text{\tiny \Eq{ell2}}}{=}\ & D\frac{\s \kappa(\s)}{\int^\s\rmd\s' \kappa(\s')}\,.\label{newds}
\ea
In the case \Eq{ell2}, the heat kernel is $Z(\s)\propto \s^{-D(1+\nu-\b)/2}$ (where the proportionality constant is the total integer volume) and
\be\label{dsnub}
\boxd{\ds=D(1+\nu-\b)\,.}
\ee
The spectral dimension depends both on the choice of weight for $\s$ and, via the parameter $\nu$, on the statistics of the underlying stochastic process. We distinguish various cases:
\begin{itemize}
\item[(i)] A natural assumption (also supported by the findings in Appendix \ref{qmtime}) is that this process is the counterpart of Brownian motion, so $\nu=1$ and $\ds=D(2-\b)$.
	\begin{itemize} 
	\item[(a)] If one further assumes that $\b=1$, one recovers the case of the old diffusion equation with nonregularized heat kernel $Z$, Sec.\ \ref{old}, where $\ds=D$. Diffusion is nonanomalous, $\langle X^2_{\textsc{fssbm}}(\s)\rangle\propto \s$.
	\item[(b)] Assuming that $\s$ inherits the measure of time coordinate, $\b=\a_0$, one has $\ds=D(2-\a_0)\geq D$ and superdiffusion. 
	\item[(c)] Assuming instead that the charge $\b$ is equal to the average fractional charge \Eq{ava}, $\b=\a$, $\ds=D(2-\a)\geq D$ and again one has superdiffusion. 
  \end{itemize}
\item[(ii)] The evolution parameter $\s$ does not have to inherit the measure of time coordinate, and there is no compelling reason why $\b$ should coincide with the fractional charges $\a_0$ or $\a$, as in cases (b) and (c). One could set $\b=1$ from the start, as in the old diffusion equation, but reintroduce $\a$ dependence ``from the backdoor,'' via a scaled Brownian motion $X=Y$ with $\nu=\a$. Then $\ds=D\a$ as in the old diffusion equation case with regularized $\cP$.
\end{itemize}
None of these cases corresponds to a fractal in the usual sense, since from Eq.\ \Eq{msdn} one sees that $\langle X^2_{\textsc{fssbm}}(\s)\rangle\propto \s^{\ds/D}\neq \s^{\ds/\dh}$. The origin of the violation of relation \Eq{dw} can be understood by revisiting the argument in Sec.~\ref{stsd} with the form of the density of states found in Sec.~\ref{old}. Apart from the replacement $\rho(E)\to w(E) \rho(E)$ (which, in fact, is only a redefinition of the density and one can just put $w=1$), the main difference is in the state counting per Hausdorff volume, which is not proportional to $\rmd\vr(x)$ but to $\rmd^Dx$. Thus, one should replace $\dh$ in Eqs.\ \Eq{E1} and \Eq{rhoed} with the topological dimension $D$. Thus, for this model of fractional spacetimes the relation \Eq{dw} is replaced by
\be
\dw=2\frac{D}{\ds}\,.
\ee

We conclude with a remark on the various choices (i)--(ii). The diffusion equation is derived, or assumed, in a classical-mechanics context. Here, the only information about an anomalous dimension is given by the measure in the time direction, while the only information about spatial dimensions is the number $D-1$ of particles $x^i(t)$, which is simply the topological dimension of space. From the point of view of the diffusion equation, the time direction is an external, arbitrary parameter $\s$. There is an intrinsic element of ambiguity in the whole construction which can be removed only by a \emph{definition} of the statistics of the random noise. Depending on which choice is regarded as ``fundamental,'' the final output will be different. The problem in this model (and the next) of fractional spaces is that none of the above choices seems to be well motivated when looking at the analogous procedure in ordinary Euclidean space. In Sec.\ \ref{qt} we shall consider another model where this conundrum is apparently solved.


\subsection{Ordinary Laplacian}\label{ordi}

Equation \Eq{tddcf} is not self-adjoint, which is a rather common situation for general Fokker--Planck equations \cite{Zwa01}. However, in a quantum-gravity setting, or in any case where the diffusion equation is aimed at the determination of the spectral dimension of spacetime, one may wonder whether the physical and geometric consequences of the diffusion equation change if one takes its adjoint. In the case of \Eq{tddcf}, $\cK_v^\dagger$ is simply the ordinary Laplacian and the adjoint equation is
\bs\label{tddcf2}\ba
&& [\p_\s-\kappa(\s)\N^2_x]\tilde P_{\b,\nu}(x,x',\s)=0\,,\\
&& \tilde P_{\b,\nu}(x,x',0)=\de_v(x,x')\,.
\ea\es
This diffusion equation actually corresponds to the scenario with standard integer Laplacian. In that case, one can show that $\b$ is the exponent governing the power law of the quantum-mechanical time (Appendix \ref{qmtime}). 

The associated stochastic process is the scaled Brownian motion
\be\label{opsbm}
\boxd{X(\s)=X_{\textsc{fsbm}}(\s) := X_{\textsc{bm}}(\s^{1+\nu-\b})\,.}
\ee
In fact, in order to construct the GLE associated with this spacetime, one should account for both the differential structure (ordinary derivatives) and the fractional generalization of the source $\xi$. The first information suggests that the left-hand side of the Langevin equation should be of the form $\p_\s X$. The second information yields, in the simplest case $\nu=1$, $\langle \xi^2\rangle\propto\de_v(\s,\s')$ and $\xi(\s)\propto \eta(\s)/\sqrt{v(\s)}$, where $\eta$ is a white noise. The result for general $\nu$ is then $\p_\s X = \s^{(\nu-\b)/2}\eta$ for the weight \Eq{vs}, which is \Eq{gleb} with $\nu$ replaced by $1+\nu-\b$. The argument is somewhat heuristic, since ordinary derivatives mix with weights in this model upon integrating by parts, and it is not obvious what a complete GLE should look like. Anyway, it yields the diffusion equation \Eq{tddcf2} and the same anomalous scaling \Eq{fssbmv} of the previous model. A trajectory of the walker is shown in Fig.\ \ref{fig2}.

The solution of Eq.\ \Eq{tddcf2} is again a Gaussian like \Eq{unb},
\be
\tilde P_{\b,\nu}=C(x',\s) \exp\left[\frac{-|x-x'|^2}{4\ell^2(\s)}\right],
\ee
but with a nonstandard normalization $C$ depending on the initial point $x'$. This is due to the initial condition in \Eq{tddcf2}, different with respect to that of Eq.\ \Eq{tddc}. We impose
\be\label{norc}
1=C(x',\s)\int_{-\infty}^{+\infty}\rmd^D x\, v(x)\, \rme^{-\frac{|x-x'|^2}{4\ell^2(\s)}}\,.
\ee
Consider the case of fixed dimensionality, $v=v_\a$. The integral factorizes and can be done analytically for each direction (\cite{GR}, formul\ae\ 3.462.1 and 9.240). Inverting the final result yields the normalization constant 
\ba
C(x',\s) &=& \prod_\mu\left\{\frac{\Gamma(\a_\mu/2)}{\Gamma(\a_\mu)}\,[2\ell(\s)]^{\a_\mu}\right.\nonumber\\
&&\qquad\left.\times\Phi\left[\frac{1-\a_\mu}2;\frac12;-\frac{|{x'}^\mu|^2}{4\ell^2(\s)}\right]\right\}^{-1}\!,
\ea
where $\Phi$ (also called $_1F_1$ or $M$) is Kummer's confluent hypergeometric function of the first kind:
\be\label{kumm}
\Phi(a;b;\,z):= \sum_{n=0}^{+\infty} \frac{(a)_n}{(b)_n} \frac{z^n}{n!}\,,
\ee
where $(a)_{n}=\Gamma(a+n)/\Gamma(a)$ is the Pochhammer symbol. To check that the initial condition in \Eq{tddcf2} is respected, we recall the asymptotic limit
\be\label{largez}
\Phi(a;b;z)\ \stackrel{\text{\tiny $z\to-\infty$}}{\sim}\ \frac{\Gamma(b)}{\Gamma(b-a)}(-z)^{-a}\,.
\ee
Thus,
\ba
C(x',\s) &\ \stackrel{\text{\tiny $\s\to 0$}}{\sim}\ &\prod_\mu\left[\frac{\Gamma(1/2)}{\Gamma(\a_\mu)}\ell^{\a_\mu} \left|\frac{{x'}^\mu}{\ell}\right|^{\a_\mu-1}\right]^{-1}\nonumber\\
&=&\frac{1}{(4\pi\ell^2)^{D/2}}\frac{1}{v_\a(x')}\,,\label{largeC}
\ea
and one recovers the fractional delta distribution $\de_v$.

Next, we calculate the asymptotic limits of the heat kernel
\ba
Z(\s)&=&\int\rmd^Dx\, v_\a(x)\,\tilde P_{\b,\nu}(x,x,\s)\nonumber\\
&=&\int\rmd^Dx\, v_\a(x)\,C(x,\s)\,.
\ea
This expression is not a power law in $\s$, so in general it will not give a constant spectral dimension as in the adjoint case. We estimate $\ds$ at small diffusion scales, Eq.\ \Eq{dsdef3}. Using again \Eq{largeC}, we get $Z(\s)\sim \cV(4\pi\ell^2)^{-D/2}$, where $\cV=\int\rmd^Dx$ is the ordinary volume. Thus, one recovers Eq.\ \Eq{dsnub}.

For large $\s$, $\Phi\to 1$, $C(x,\s)\sim \prod_\mu[\Gamma(\a_\mu/2)/\Gamma(\a_\mu)$ $\ell^{\a_\mu}(\s)]^{-1}$, $Z(\s)\propto \cV_{\rm H} \ell^{-D\a}(\s)\propto \s^{-D\a(1+\nu-\b)/2}$, and the spectral dimension is $\a$ times smaller than the one at small scales. This mismatch can be explained as an artifact of adopting a measure corresponding to a geometry with fixed dimensionality. Characteristic scales appear either as topological effects (as, for instance, the curvature radius of a sphere or a torus) or in dimensional flows determined by an intrinsically multiscale measure. Here there is no foothold to establish a scale hierarchy, so it is more natural to take Eq.\ \Eq{dsdef3} as the correct definition of $\ds$ when characteristic scales are not expected.  Therefore, the limit $\s\to\infty$ of the return probability has no significance here and the result for $\ds$ can be interpreted as in agreement with the one from the adjoint diffusion equation. Further support for this conclusion will be given in Sec.\ \ref{muso}, where we shall compute $\ds$ in the multiscale scenarios. In the more realistic case where $\ds$ changes with the scale, the spectral dimension of the two models is in perfect agreement.


\subsection{\texorpdfstring{$q$}{}-Laplacian}\label{qt}

\begin{figure}
\centering
\includegraphics[width=8.4cm]{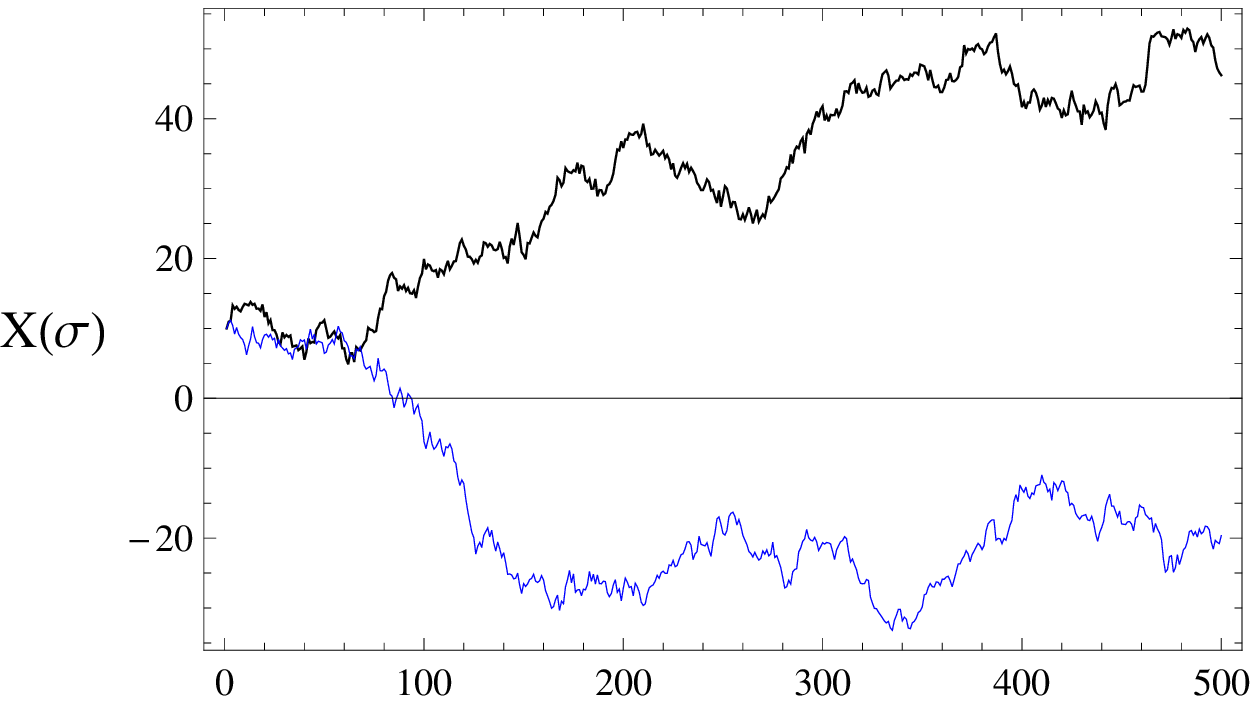}\\
\includegraphics[width=8.4cm]{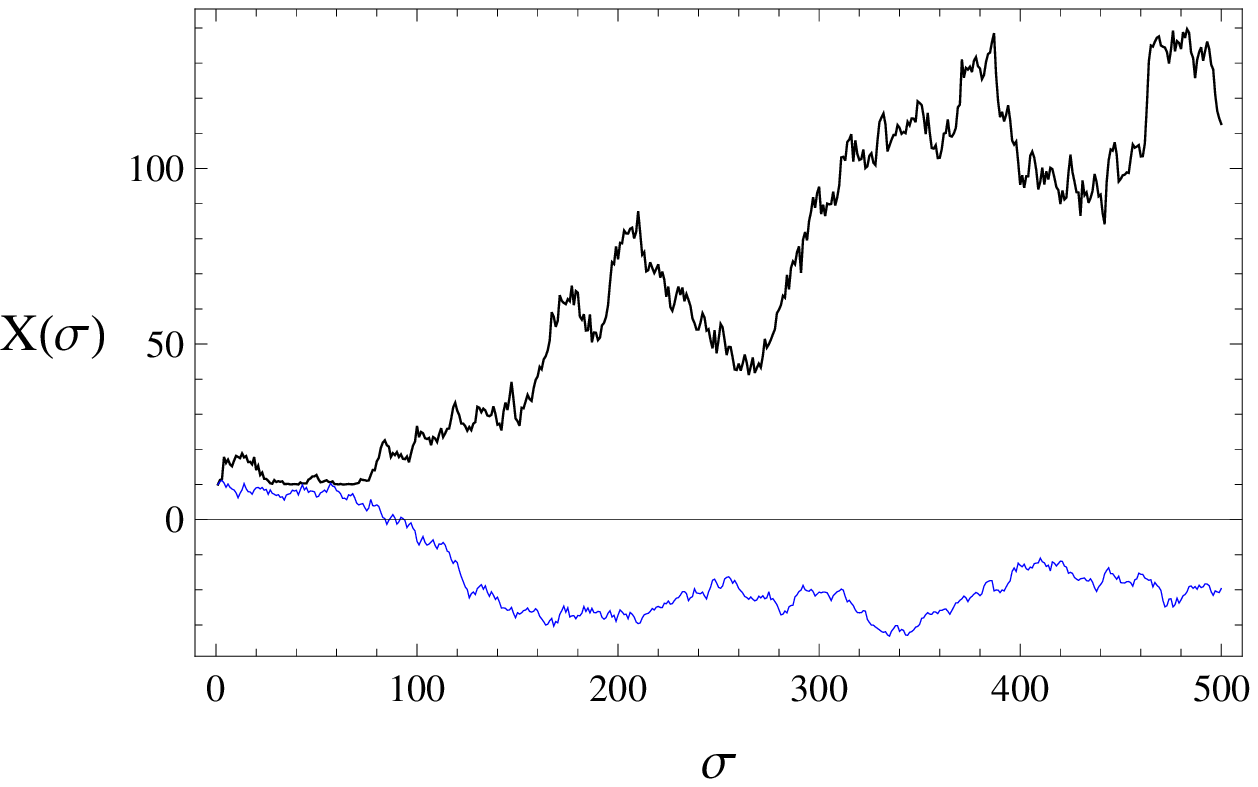}
\caption{\label{fig3} Dark (black) curve: Example of trajectory of the process \Eq{fsbmq} for the diffusion measure weight \Eq{vs}, with $\b=1=\a$ (top panel, ordinary Brownian motion in ordinary spacetime) and $\b=1/2=\a$ (bottom panel, fractional spacetime with $q$-Laplacian). Light (blue) curve: Example of trajectory of ordinary Brownian motion in ordinary spacetime, plotted for reference (notice the different scaling of the vertical axes).}
\end{figure}

This scenario has a standard classical mechanics (ordinary derivatives in a certain diffusion time $\tau$). All the results of Sec.\ \ref{euc} hold with the following changes:
\begin{enumerate}
\item[(i)] The random variable $X(\s)$ is now regarded as a composite variable $Q(\tau)=Q[X(\tau)]$. In the case of fixed dimensionality, for each direction
\be\label{QX}
Q^\mu[X(\tau)]=\frac{{\rm sgn}[X^\mu(\tau)]|X^\mu(\tau)|^{\a_\mu}}{\Gamma(\a_\mu+1)}\,.
\ee
This expression is invertible, since ${\rm sgn}(X)={\rm sgn}(Q)$ and $X={\rm sgn}(Q)[\Gamma(\a+1) |Q|]^{1/\a}$.
The parameter $\tau$ is by itself composite, $\tau=\vr(\s)$, and in particular
\be\label{taus}
\tau= \vr_\b(\s)=\frac{\s^\b}{\Gamma(\b+1)}
\ee
in the no-scale fractional case, where $\s$ has the dimensionality of a length and $\b$ is a constant. 
\item[(ii)] The diffusion equation is
\ba
&&\left\{\frac{\p}{\p\tau(\s)}-\sum_\mu \kappa_{(\mu)}\frac{\p^2}{[\p q(x^\mu)]^2}\right\}P=0\,,\label{deq}\\
&& [\tau]=-\b\,,\qquad [\kappa_{(\mu)}]=\b-2\a_\mu\,,
\ea
where, in the general anisotropic case, the diffusion coefficient is different for each direction. One can also define the parameter
\ba
&&\ell_\mu(\s):=\ell_\mu[\tau(\s)]=\sqrt{\kappa_{(\mu)} \tau(\s)}\,,\label{BMlent}\\
&& [\ell_\mu]=-\a_\mu\,.
\ea
In the isotropic case ($\a_\mu=\a$), there is only one diffusion coefficient $\kappa$ and one parameter $\ell(\s):=\sqrt{\kappa\tau(\s)}$ with scaling $[\ell]=-\a$. Then, the diffusion equation \Eq{deq} can be written as
\be\label{deq2}
\left\{\frac{\p}{\p\ell^2(\s)}-\sum_\mu \frac{\p^2}{[\p q(x^\mu)]^2}\right\}P=0\,.
\ee
This version of the diffusion equation is also valid in the anisotropic case when the dimensionful coefficients are absorbed in the definitions of $q^\mu$, in which case $\ell^2=\tau$.
\item[(iii)] Thus, as a function of the actual diffusion time $\s$ (which is the one with respect to which one must take the derivative of the heat kernel in order to get $\ds$) the random variable $Q(\s)=Q[\tau(\s)]$ is a scaled Brownian motion. 
In the particular case of Eq.\ \Eq{QX}, in the first orthant one can identify the stochastic process, which we dub FSBM-$q$, as [up to $\Or(1)$ constants]
\ba
\boxd{X_{\textsc{fsbm-}q}(\s) \sim [X_{\textsc{bm}}(\s^\b)]^{1/\a}= [X_{\textsc{sbm}}(\s)]^{1/\a}.}\nonumber\\\label{fsbmq}
\ea
The raggedness of the ordinary SBM increases with increasing $\b$ (Fig.\ \ref{fig2}). On the other hand, with respect to an ordinary SBM the raggedness of FSBM-$q$ is greater ($1/\a>1$) and increases for smaller $\a$; the drift from the average is then amplified. There is an interplay of the two effects, but the effect from $\a$ (power of $Q$) is greater than the one from $\b$ (scaled time), even when $\b=\a$. Therefore, this incarnation of FSBM is typically more ragged than an ordinary Brownian motion; see Fig.\ \ref{fig3}.
\item[(iv)] Equation \Eq{Gau} holds with the coordinate replacement $x\to q(x)$,
\be\label{Gauq}
P(x,x',\s)=\prod_\mu\frac{\rme^{-\frac{|q(x^\mu)-q({x'}^\mu)|^2}{4\ell_\mu^2(\s)}}}{\sqrt{4\pi\ell_\mu^2(\s)}}\,,
\ee
where the normalization is automatically correct.
\item[(v)] The heat kernel is of the form \Eq{Zfr}, which for the fractional case \Eq{taus} reads
\be\label{Zfr2}
Z(\s)=\frac{\cV_{\rm H}}{\prod_\mu\sqrt{4\pi\ell_\mu^2(\s)}}\propto \cV_{\rm H}\,\s^{-\frac12D\b}\,,
\ee
so that the first Seeley--DeWitt coefficient does correspond to the Hausdorff volume (this fixes a problem unsolved in the other models) and the spectral dimension is
\be\label{dsdb}
\boxd{\ds=D\b\,.}
\ee
\end{enumerate}

As in the other theory, the spectral dimension only depends on the topological dimension of space and on the fractional charge associated with diffusion time. The latter stems from a classical mechanical model which knows nothing about the Hausdorff dimension of spacetime, since it only sees $D$ motions $Q^\mu$. Again, $\s$ is just a parameter from the point of view of spacetime, and $\b$ does not necessarily have to be equal to the time fractional charge $\a_0$. To fix this arbitrariness, we can encode the information of the Hausdorff dimension in $\b$ and define it as the average fractional charge, $\b=\a=\dh/D$. This way, it results that $\ds=\dh$, as in Eq.\ \Eq{dsda}. 

Even for a general $\b$, the walk dimension is indeed the one for a fractal, since from $Q\sim \tau^{1/2}$ there follows $X\sim \s^{\b/(2\a)}$, and Eq.\ \Eq{dw} holds.


\section{Diffusion in multifractional spacetimes}\label{mus}

Assume that the distribution $v(\s)$ takes the same form of the multifractional measure weight \Eq{muf2},
\be\label{vmun}
v(\s)=v_*(\s)=\sum_{n=1}^N g_n \s^{{\b_n-1}}\,,
\ee
and for the sake of simplicity consider the binomial case $g_1=\ell_*^{1-\b_*}$, $0<\b_1=\b_*<1$, $g_2=1$, and $\b_2=1$:
\be\label{vmu}
v_*(\s)=1+\left(\frac{\s}{\ell_*}\right)^{{\b_*-1}},
\ee
where $\ell_*$ is a fundamental length discriminating between infrared ($\s\gg \ell_*$) and ultraviolet ($\s\ll \ell_*$). Also fix $\nu=1$ in the statistics of the random walker and call $\kappa_{\b,1}=\kappa_\b$.


\subsection{Weighted Laplacian}

When the spacetime measure weight $v(x)$ encodes a multiscale geometry, the theory with weighted Laplacian does not fare well, although it can be rescued by an interesting modification of the physical interpretation. 

The multiscale version of the process \Eq{fsbm} is obtained by using the distribution \Eq{vmun} or \Eq{vmu}; we do not plot the resulting trajectories here. From now on we aim to compute the spectral dimension.

Let $1<\b_*<2$. From the asymptotic behaviour of \Eq{vmu}, one expects that in the small-scale limit the first term dominates and $\ds\sim D$, while in the large-scale limit the second term takes the lead and one recovers $\ds\sim D(2-\b_*)\leq D$, i.e., a reduction of the spectral dimension with the scale. This is confirmed by an explicit calculation. Plugging \Eq{vmu} in Eq.\ \Eq{ell2}, we obtain
\ba
\ell^2(\s)&=&\bar{\ell}^2+\int^\s\frac{\rmd\s'}{1+(\s'/\ell_*)^{\b_*-1}}\nonumber\\
&=&\bar{\ell}^2+\kappa_{\b}\s\, F\left[1,\frac{1}{\b_*-1};\frac{\b_*}{\b_*-1};-\left(\frac{\s}{\ell_*}\right)^{\b_*-1}\right],\nonumber\\\label{ell2F}
\ea
where we used formula 3.194.5 of \cite{GR} 
 and $F={}_2F_1$ is the hypergeometric function
\be
F(a,b;c;\,z):= \sum_{n=0}^{+\infty} \frac{(a)_n(b)_n}{(c)_n} \frac{z^n}{n!}\,.
\ee
For $\s\ll\ell_*$, $F\to 1$ and $\ell^2(\s)\to \bar{\ell}^2+\kappa_{\b}\s$, thus fixing $\bar{\ell}=0$. For $z\to\infty$ one must analytically continue via formula 9.132.1 of \cite{GR}: 
\ba
 F(1,b;b+1;\,z)&=&\frac{b}{b-1}\frac{1}{1-z} F\left(1,1;2-b;\frac{1}{1-z}\right)\nonumber\\
&&+\Gamma(b+1)\Gamma(1-b)\frac{1}{(-z)^b}\,.\label{ff}
\ea
For $\b_*>1$, $b=1/(\b_*-1)>1$, so that
\be\label{ellpeq1}
\ell^2(\s)\sim \begin{cases} \kappa_{\b}\s & \text{($\s\ll\ell_*$)}\\
\frac{\kappa_{\b}\s}{2-\b_*}\left(\frac{\s}{\ell_*}\right)^{1-\b_*} & \text{($\s\gg\ell_*$)} \end{cases}\,,\quad \text{for $1<\b_*<2$}\,.
\ee
From Eq.\ \Eq{newds}, the spectral dimension reads
\be
\ds(\s)=\frac{D\kappa_{\b}\s}{v_*(\s)\ell^2(\s)}\,.\label{newdsmf1}
\ee
From the asymptotic behaviour \Eq{ellpeq1}, it immediately follows that
\be\label{dsasi}
\ds\sim \begin{cases} D & \text{($\s\ll\ell_*$)}\\
D(2-\b_*)\leq D & \text{($\s\gg\ell_*$)} \end{cases}\,,\quad \text{for $1<\b_*<2$}\,,
\ee
as announced.

Let us now consider the case $0<\b_*<1$ ($b<-1$). Again, one expects a decrease of the spectral dimension as the scale increases, from $\ds\sim D(2-\b_*)\geq D$ to $\ds\sim D$. Equation \Eq{ell2F} is valid only for $\b_*>1$, but we can analytically continue it under the provision that $\ell^2(0)=0$. At large scales, $\ell^2(\s)\sim \bar{\ell}^2+\kappa_{\b}\s$, while for $\s\gg\ell_*$ we use Eq.\ \Eq{ff}, this time picking both terms. Noting that $\Gamma(b)\Gamma(1-b)=\pi/\sin(\pi b)$, from Eq.\ \Eq{ell2F} we get
\be\label{ell*}
\ell^2(\s)\ \stackrel{\text{\tiny $\s\ll\ell_*$}}{\sim}\ \left[\bar{\ell}^2 +\kappa_{\b}\ell_* \frac{\frac{\pi}{\b_*-1}}{\sin \frac{\pi}{\b_*-1}}\right]+\frac{\kappa_{\b}\s}{2-\b_*}\left(\frac{\s}{\ell_*}\right)^{1-\b_*},
\ee
implying that
\be
\bar{\ell}^2=-\kappa_{\b}\ell_* \frac{\frac{\pi}{\b_*-1}}{\sin \frac{\pi}{\b_*-1}}\,.
\ee
This formula is valid also for $\b_*=1+1/k$, $k\in\mathbbm{Z}$, where $\bar{\ell}^2\to-\kappa_{\b}\ell_*$. To summarize,
\be\label{ellpeq2}
\ell^2(\s)\sim \begin{cases} 
\frac{\kappa_\b\s}{2-\b_*}\left(\frac{\s}{\ell_*}\right)^{1-\b_*} & \text{($\s\ll\ell_*$)}\\
\bar{\ell}^2+\kappa_{\b}\s & \text{($\s\gg\ell_*$)}\end{cases}\,,\quad \text{for $0<\b_*<1$}\,,
\ee
and we get an interchange of the regimes of \Eq{dsasi}:
\be\label{dsasi2}
\ds\sim \begin{cases} 
D(2-\b_*)\geq D & \text{($\s\ll\ell_*$)}\\
D & \text{($\s\gg\ell_*$)} \end{cases}\,,\quad \text{for $0<\b_*<1$}\,,
\ee
although the decreasing behaviour remains.

The overall scenario is radically different from the ``embedding picture'' previously associated with models with weighted Laplacians. In that case, one started with a $D$-dimensional ambient Minkowski spacetime $M^D$ and had a multifractional spacetime $\cM_*$ embedded in it. Then, $\cM_*$ ``filled'' $M^D$ only at large scales. This is true also in the present case, but only for the Hausdorff dimension. On the contrary, there is a reduction of the spectral dimension at large scales:
\begin{enumerate}
\item[I.] For $0<\b_*<1$ [Eq.\ \Eq{dsasi2}], it is as if fractional spacetime gushed out of the embedding at smaller and smaller scales. Taking, for instance, $D=4$ and $\b=1/4$, $1/2$, $3/4$, one would have a dimensional reduction $\ds\sim 7$, $6$, $5\to \ds\sim 4$.
\item[II.] For $1<\b_*<2$ [Eq.\ \Eq{dsasi}], it is as if the embedding were depleted of spacetime points at larger and larger scales. For example, for $D\geq 5$ and $\b=2-4/D$, one gets $\ds\sim 4$ in the infrared starting from an embedding with five or more dimensions.
\end{enumerate}
Possible modifications may come from taking a more general statistics for the diffusion process ($\nu\neq 1$, but this is not natural from the point of view of quantum mechanics; see the appendices) or from changing the initial condition from a delta to a Gaussian. As discussed in Sec.\ \ref{abm}, this can be achieved simply by letting the constant $\bar{\ell}^2$  be nonvanishing by default. The interpretation then would be the one of \cite{SmS1,SmS2,SmS3,MoN}, namely, that the test particle is not pointwise due to the intrinsic ``fuzziness'' of a spacetime with a minimal length. To get this type of structure in the multifractional context, it would be quite natural to identify the characteristic scale $\ell_*$ in the binomial measure with the scale at which fuzziness effects become important. In turn, this univocally sets
\be
\bar{\ell}=\ell_*\,.
\ee
This would ``straighten'' dimensional flow. Take, in fact, $0<\b_*<1$. The only but crucial modification to the above calculation of $\ds$ is that the constant term in Eq.~\Eq{ell*} does \emph{not} vanish. Call $L^2$ the content in square brackets. While the large-scale limit of the spectral dimension is unaffected, the small-scale one is
\be
\ds \stackrel{\text{\tiny $\s\ll\ell_*$}}{\sim} D \frac{\kappa_{\b}\ell_*}{L^2}\left(\frac{\s}{\ell_*}\right)^{2-\b_*},\quad L^2=\bar{\ell}^2 +\kappa_{\b}\ell_* \frac{\frac{\pi}{\b_*-1}}{\sin \frac{\pi}{\b_*-1}}\,,
\ee
leading to
\be\label{dsasi3}
\ds\sim \begin{cases} 0 & \text{($\s\ll\ell_*$)}\\
D & \text{($\s\gg\ell_*$)} \end{cases}\,.
\ee
Such a scenario somewhat merges multifractal\footnote{Here we can talk about fractals because $\ds\leq \dh$.} and effective noncommutative geometries, but in a way different from \cite{ACOS}. In \cite{ACOS}, $\kappa$-Minkowski spacetime was reinterpreted as the ultramicroscopic limit of a log-oscillating fractional measure. On the other hand, here we have changed the initial condition of diffusion (but, implicitly, also of the Green function) so as to have the scale in the real-order multifractional measure play the role of a fuzziness length, just as in commutative effective spacetimes coming from a noncommutative geometry where operators are evaluated on coherent states \cite{SmS1,SmS2,SmS3}. However, the matching is only at the level of the spectral dimension \cite{MoN}, since our resolution of the identity is not a Gaussian but a fractional delta. We leave a further elaboration of this scenario and a fuller assessment of its physical consequences for future study.


\subsection{Ordinary Laplacian}\label{muso}

The normalization condition \Eq{norc} now features a multiscale measure weight $v(x)=v_*(x)$. With the binomial measure \Eq{mufbin} and in the isotropic case $\a_\mu=\a_*$, one has
\ba
 C(x',\s) &=&\left\{[4\pi\ell^2(\s)]^{\frac{D}2}+\ell_*^D\left[\frac{\Gamma(\a_*/2)}{\Gamma(\a_*)}\,\frac{2^{\a_*}\ell^{\a_*}(\s)}{\ell_*^{\a_*}}\right]^D\right.\nonumber\\
&&\quad\left.\times\prod_\mu\Phi\left[\frac{1-\a_*}2;\frac12;-\frac{({x'}^\mu)^2}{4\ell^2(\s)}\right]\right\}^{-1}\!,
\ea
where $\ell^2(\s)$ has been computed in the previous subsection. As before, we focus on the asymptotic limits of the heat kernel $Z(\s)=\int\rmd^Dx\, v_*(x)\,C(x,\s)$ and of the spectral dimension. For $0<\b_*<2$, $\ell^2\to 0$ when $\s\to 0$, and the initial condition is recovered, since from Eq.\ \Eq{largez}
\be\label{largeC2}
C(x',\s)\ \stackrel{\text{\tiny $\s\to 0$}}{\sim}\ \left[(4\pi\ell^2)^{\frac{D}2}v_*(x')\right]^{-1}.
\ee
Thus, $Z\sim [\ell(\s)]^{-D}$. Still for the whole range $0<\b_*<2$, when $\s\to+\infty$ one has $\ell^2\to+\infty$, $\Phi\to 1$, and again $Z\sim [\ell(\s)]^{-D}$. From Eqs.\ \Eq{ellpeq1} and \Eq{ellpeq2}, one eventually gets Eqs.\ \Eq{dsasi} and \Eq{dsasi2}. 

This result amends the incomplete discussion in Sec.\ \ref{ordi}. With respect to dimensional flow, this model falls into the same class of its Hermitian-conjugate dual, i.e., the model with weighted Laplacian.


\subsection{\texorpdfstring{$q$}{}-Laplacian}

The case of $q$-theory with scale-dependent dimension is straightforward. One should simply replace $\ell^2(\s)$ in Eq.~\Eq{deq2} and the distributions $q(x^\mu)$ with a generic functional of, respectively, $\s$ and $x^\mu$. The multifractional measure weight \Eq{muf2} determines the multiscale form of $q$:
\be\label{qx}
q(x^\mu)=\vr_*(x^\mu) =\sum_{n=1}^N g_{\mu,n} {\rm sgn}(x^\mu)|x^\mu|^{\a_\mu}\,,
\ee
where we absorbed $\Gamma$ factors into the $g_{\mu,n}$'s. This also suggests the form of $\ell^2$:
\be\label{ell2mf}
\ell^2(\s)=\vr_*(\s)=\sum_{n=1}^N g_n \s^{\b_n}\,.
\ee
The simplest case is the binomial measure \Eq{vmu},
\be\label{ell2mf2}
\ell^2(\s)=\kappa\ell_*\left[\frac{\s}{\ell_*}+\left(\frac{\s}{\ell_*}\right)^{\b_*}\right]\,.
\ee
Since $0<\b_*<1$ in this model, at small scales $\s/\ell_*\ll 1$ the second term dominates and $\ell^2\propto\s^{\b_*}$, while at large scales $\s/\ell_*\gg 1$ the variance of the PDF is the usual one, $\ell^2=\kappa\s$.

In general, the coefficients $g_n$ may also depend on $\s$, according to the type of profile one wishes to describe. Barring this possibility, from Eqs.\ \Eq{deq2}, \Eq{ell2mf}, and \Eq{newdsq}, the spectral dimension reads
\be
\boxd{\ds(\s) = D\frac{\sum_n g_n \b_n \s^{\b_n}}{\sum_n g_n \s^{\b_n}}\,.}
\ee
In particular, for the case of fixed dimensionality ($N=1$, $\b_1=\b$), the spectral dimension coincides with Eq.\ \Eq{dsdb}, while for the binomial measure one has
\be
\ds(\s) = D\frac{1+ \b_* (\ell_*/\s)^{1-\b_*}}{1+ (\ell_*/\s)^{1-\b_*}}\,.
\ee
As anticipated, $\ds\sim D\b_*$ at small scales $\s/\ell_*\ll 1$ and $\ds\sim D$ at large scales $\s/\ell_*\gg 1$.

Since Eq.\ \Eq{qx} is not invertible, knowing the statistics of $Q$ does not immediately lead to a statistics for $X$. One can, however, apply Eq.\ \Eq{fsbmq} at various asymptotic regimes with different $\a$, and give a different, patchwise stochastic description of spacetime for each scale range where $\ds\sim {\rm const}$.


\section{Discussion}\label{disc}

The three classes of no-scale spacetime models analyzed here display constant anomalous dimensions and are associated with highly nontrivial stochastic processes, all related (via various rescalings and mappings) to Brownian motion or scaled Brownian motion. Table \ref{tab1} summarizes these findings. To the best of our knowledge, these processes do not have counterparts in the literature of probability theory and we had to name them according to the spacetime they describe.
\begin{table*}
\begin{ruledtabular}
\caption{\label{tab1} Models of fractional spacetimes with fixed dimensionality and $\nu=1$, characterized by the symmetries of the Lagrangian density, a stochastic process $X(\s)$, Hausdorff dimension $\dh$ and spectral dimension $\ds$.}
\centering
\begin{tabular}{lcccc}
Model  (symmetries of $\cL$) & d'Alembertian & Stochastic process & $\dh$ & $\ds$ \\ \hline
Weighted Poincar\'e & $\frac{1}{\sqrt{v}}\B_x (\sqrt{v}\,\cdot\,)$ Eq.\ \Eq{ka} & $X_{\textsc{fsbm-}v}(\s)$ Eq.\ \Eq{fsbm} & $D\a$ Eq.\ \Eq{had} & $D(2-\b)$ Eq.\ \Eq{dsnub} \\
Ordinary Poincar\'e &   $\B_x$,~~ $\B_x^\dagger=\frac{1}{v}\B_x (v\,\cdot\,)$ Eqs.\ \Eq{ka1} and \Eq{boda} & $X_{\textsc{fsbm}}(\s)$ Eq.\ \Eq{opsbm} & $D\a$ & $D(2-\b)$ \\
$q$-Poincar\'e & $\B_{q(x)}$ Eq.\ \Eq{boq} & $X_{\textsc{fsbm-}q}(\s)$ Eq.\ \Eq{fsbmq} & $D\a$ & $D\b$ Eq.\ \Eq{dsdb} \\
\end{tabular}
\end{ruledtabular}
\end{table*}

We also got the analytic expression of the spectral dimension of the multiscale extensions of these theories, found that the models with weighted and ordinary Laplacian (which are Hermitian conjugate) produce the same dimensional flow, and gave an alternative interpretation of the theory with weighted Laplacian as a fuzzy spacetime. In this case, dimensional flow is modified accordingly and $\ds\to 0$ in the UV.

Only the model with $q$-Laplacian is fractal in the usual sense. Namely, the first Seeley--DeWitt coefficient in the heat kernel expansion corresponds to the Hausdorff volume and the relation among Hausdorff, spectral, and walk dimensions is \Eq{dw}. The other models violate these conditions, mainly because the effective density of states scales as the integer embedding volume. Such properties are likely to signal a difference in the renormalization of field theories living on these spacetimes.

In all the models of fractional and multiscale spacetimes we have considered in this paper, the spectral dimension was derived starting from statistical mechanics, via a generalized Langevin equation. There is, in general, an intrinsic ambiguity in the diffusion equation approach, inasmuch as it does not fix the scaling dimension of effective diffusion time. This, in fact, is part of the definition of the theory, and one should not expect to get a certain value of $\ds$ without imposing a certain number of defining conditions. Part of this ambiguity might be related to other aspects of the theory which we have not considered here, such as the momentum-space structure. Taking as a guiding principle the conjecture \cite{Akk2} according to which the spectral dimension $\ds$ is the dimension of momentum space, and fixing the latter by assuming that momentum transform is an automorphism [hence, $w(k)=v(k)$], one would fix the value of $\b$ \emph{a posteriori}. Similar considerations are not rigorous at the present stage and further study will be needed.

The above approach is classical. Quantum mechanics can provide valuable information about the diffusion equation, which, as a matter of fact, can be derived as the classical limit of a quantum diffusion process. In the appendices we discuss some aspects of the relation between classical and quantum diffusion.


\bigskip

\begin{acknowledgments}
The work of G.C.\ is under a Ram\'on y Cajal tenure-track contract.
\end{acknowledgments}


\begin{appendix}

\section{QUANTUM PROBABILITY DENSITY AS A BILINEAR}\label{qmbil} 

In finding the diffusion equation \Eq{tddcf} we did not give a robust motivation to the form of the Laplacian $\check{\cK}_v$, i.e., of the functional form \Eq{Puv}. We can do so by a simple calculation of the quantum probability density function in multiscale spacetimes with weighted Poincar\'e symmetries, closely following the same procedure as in ordinary spacetime \cite{AkM}. Since we are in a quantum-mechanical context, we denote time with $t$. We also write $D$-dimensional spatial vectors in normal font.

Consider the quantum-mechanical free particle with unit mass and Hamiltonian $\hat H_{\rm free}=-\cK_v^{\rm E}$, where $\cK_v^{\rm E}$ is the Euclidean version of \Eq{ka} in $D$ spatial dimensions. The energy $E_k=k^2$ is given by the eigenvalue equation $\hat H_{\rm free}\bE(k,x)=E_k\bE(k,x)$, where we used the ``plane waves'' \Eq{E}. The associated Green equation with delta source both in space and time ($\hbar=1$) is
\be\label{gequ}
(\rmi\cD_t-\hat H_{\rm free})G_{\rm free}(x,t;x',t')=\de_{v_0}(t,t')\,\de_v(x,x')\,.
\ee
The solutions are the advanced and retarded propagators
\be
G^{{\rm R},{\rm A}}_{\rm free}(x,t;x',t')=\mp\rmi\frac{\theta[\pm(t-t')]}{\sqrt{v_0(t)v_0(t')}}\,\langle x|\rme^{-\rmi \hat H_{\rm free} (t-t')}|x'\rangle\,,
\ee
where $\theta$ is the Heaviside step function. Transforming into energy-momentum space, and using
\be\nonumber
\mp\rmi\theta(\pm z)=\int_{-\infty}^{+\infty}\rmd\om\,\frac{\rme^{\rmi\om z}}{\om\pm\rmi\e}\,,
\ee
it is easy to show that,
\ba
G^{{\rm R},{\rm A}}_{\rm free}(x,t;x',t')&=&\int\rmd E\,w(E)\,\bE^*(E,t)\bE(E,t')\nonumber\\
&&\qquad\times\tilde G^{{\rm R},{\rm A}}_{\rm free}(x,x';E)\,,\label{gg}
\ea
where we assumed a generic measure weight $w(E)$, we extended the plane waves \Eq{E} to the energy-time pair, $\bE(E,t)=\rme^{\rmi E t}/\sqrt{v_0(t) w(E)}$, and
\be
\tilde G^{{\rm R},{\rm A}}_{\rm free}(x,x';E)=\int\rmd^Dk\,w(k)\,\frac{\bE^*(k,x)\bE(k,x')}{E-E_k\pm\rmi\e}\,.
\ee
Making measure factors explicit and canceling them, we obtain that the fractional propagators $G^{{\rm R},{\rm A}}_{\rm free}$ are related to the ones in ordinary space $\bar G^{{\rm R},{\rm A}}_{\rm free}$ by
\be\label{Gfree}
G^{{\rm R},{\rm A}}_{\rm free}(x,t;x',t')=\frac{\bar G^{{\rm R},{\rm A}}_{\rm free}(x-x',t-t')}{\sqrt{v(x)v_0(t)v(x')v_0(t')}}\,.
\ee
Comparing with the propagator computed in \cite{frc5}, there is an extra prefactor $1/v_0(t')$ [there, the overall time-dependent prefactor is $\sqrt{v_0(t')/v_0(t)}$], because the propagator of \cite{frc5} was a solution of the Green equation only with spatial delta source. Related to this, notice that, in the limit $t\to t'$, $G^{{\rm R},{\rm A}}_{\rm free}(x,t;x',t)=\de_v(x,x')/v_0(t)$. This suggests, as a feedback, that we define the Green equation \Eq{gequ} with an ordinary time direction (ordinary delta and time derivative); its solutions are then $\sqrt{v_0(t')v_0(t)}G^{{\rm R},{\rm A}}_{\rm free}$.

Let us now turn to a generic quantum system without specifying the Hamiltonian $\hat H$ and the dispersion relation between the energy $E_k$ and momentum. We only assume that $\hat H$ is characterized by energy eigenvalues $E_k$ and eigenstates $|\psi_k\rangle$: $\hat H|\psi_k\rangle=E_k|\psi_k\rangle$. Here the momentum $k$ can be either discrete \cite{AkM} or continuous, but for simplicity we stick with the continuum case. Let $|\Psi_{x'}\rangle$ be a normalizable wave packet of average energy $\cE$ centered at the spatial point $x'$ and with dispersion $s$:
\be
|\Psi_{x'}\rangle:=A\int\rmd^D k\, w(k)\, \langle \psi_k|x'\rangle\, \rme^{-\frac{(E_k-\cE)^2}{4s}}|\psi_k\rangle\,,
\ee
where $A=A(x',s)$ is a normalization constant and we assumed a Gaussian form for the packet. Projecting on the position-space basis and defining $\psi_k(x):=\langle x|\psi_k\rangle$, one obtains the wave function
\ba
\Psi_{x'}(x) := \langle x|\Psi_{x'}\rangle=A\!\int\rmd^D k w(k)\psi_k^*(x')\psi_k(x) \rme^{-\frac{(E_k-\cE)^2}{4s}}.\nonumber\\
\ea
Scalar products of wave functions include a nontrivial weight factor, so that the orthonormality of the $\psi_k(x)$ is
\be
\langle \psi_{k'}|\psi_k\rangle:=\int\rmd^Dx\,v(x)\,\psi_k(x)\psi_{k'}^*(x)=\de_w(k,k')\,.
\ee
This relation is the starting point for finding the normalization constant $A$ by imposing $\langle \Psi_{x'}|\Psi_{x'}\rangle=1$. Defining the density of states \Eq{rhoE}, one has
\be
1=A^2\int\rmd E\, w(E) \rho_{x'}(E)\,\rme^{-\frac{(E-\cE)^2}{2s}}\,,
\ee
where we made explicit the dependence of $\rho$ on $x'$ via a subscript. The result for $A$ depends on the way the density of states is approximated. For instance, since $\rho_{x'}(E)\propto |\psi_k(x')|^2\propto [v(x')]^{-1}$,
it is natural to replace $\rho_{x'}(E)$ with $\rho_0/v(x')$, where $\rho_0$ is the constant average over disorder per integer volume \cite{AkM}. Thus, we obtain
\be\label{A2}
A^2= \frac{v(x')}{\rho_0}\,\cC(s,\cE)\,,
\ee
where the function $\cC$ depends on the form of $w(E)$ and, in fractional spaces, also on the average energy $\cE$. The details of $\cC$ are not important here, since we wish to extract only the measure-weight dependence of the quantum PDF.

The latter is defined as the square of the matrix element of the evolution operator
\ba
\hspace{-0.8cm}\cG^{\rm R}(x,t;x',t') &:=&-\rmi\frac{\theta(t-t')}{\sqrt{v_0(t)v_0(t')}}\langle x|\rme^{-\rmi \hat H(t-t')}|\Psi_{x'}\rangle\nonumber\\
&=& \int\rmd^D y\,v(y)\,G^{\rm R}(x,t;y,t')\,\Psi_{x'}(y),
\ea
averaged over disorder. The Hamiltonian, in fact, may be quite complicated by the interaction of the particle with the underlying medium. This interaction is of stochastic type and may be included as an effective noise potential. We denote this stochastic average as $\llangle\cdot\rrangle$. Thus, the probability density function of \emph{quantum} diffusion is
\begin{widetext}
\ba
\hspace{-0.8cm}P_{\rm q}(x,t;x',t') &:=& \llangle |\cG^{\rm R}(x,t;x',t')|^2\rrangle\nonumber\\
&=& A^2 \int \rmd E\, w(E)\,\bE^*(E,t)\bE(E,t') \int \rmd E'\, w(E')\,\bE^*(E',t')\bE(E',t) \,\rme^{-\frac{(E-\cE)^2+(E'-\cE)^2}{4s}}\nonumber\\
&&\qquad\times \llangle \tilde G^{\rm R}(x,x'; E)\tilde G^{\rm A}(x',x; E')\rrangle\nonumber\\
&=& \frac{A^2}{v_0(t)v_0(t')} \int \rmd E\int \rmd E'\,\rme^{-\rmi (E-E')(t-t')}\rme^{-\frac{(E-\cE)^2+(E'-\cE)^2}{4s}}\, \llangle \tilde G^{\rm R}(x,x'; E)\tilde G^{\rm A}(x',x; E')\rrangle\nonumber\\
& \stackrel{\text{\tiny $\om=E-E'$}}{=} & \frac{A^2}{v_0(t)v_0(t')} \int \rmd E\int \rmd\om\,\rme^{-\rmi \om(t-t')}\rme^{-\frac{(E-\cE)^2+(E-\om-\cE)^2}{4s}}\, \llangle \tilde G^{\rm R}(x,x'; E)\tilde G^{\rm A}(x',x; E-\om)\rrangle\,,\label{picu}
\ea
\end{widetext}
where we used the analogue of Eq.\ \Eq{gg} for the interacting case. At this point, we make two assumptions \cite{AkM}: first, that $\om\ll s$ (spread much larger than the frequency) and, second, that
\be\label{Pq1}
P_{\rm q}(x,x';\om):=\frac{1}{\rho_0}\llangle\tilde G^{\rm R}(x,x'; E)\tilde G^{\rm A}(x',x; E-\om)\rrangle
\ee
depends on the energy $E$ only weakly. We can thus take the stochastic average outside of the integral in $E$. Then, Eqs.\ \Eq{A2} and \Eq{picu} yield
\be\label{Pq2}
P_{\rm q}(x,t;x',t') \approx \frac{v(x')}{v_0(t')v_0(t)}\int\rmd\om\, \rme^{-\rmi\om(t-t')}P_{\rm q}(x,x';\om)\,.
\ee
The extra prefactor $1/[v_0(t')v_0(t)]$ would disappear if one assumed a Green equation with ordinary time delta and time derivative instead of \Eq{gequ}.

From Eqs.\ \Eq{Gfree}, \Eq{Pq1}, and \Eq{Pq2}, it follows that the coordinate dependence of the quantum PDF in fractional spaces is the usual one times a prefactor (spatial vector notation reinstated) $v({\bf x}')[v({\bf x}')v({\bf x})v_0(t') v_0(t)]^{-1}=[v({\bf x})v_0(t)]^{-1} [v_0(t')]^{-1}\propto[v(x)]^{-1}$. The \emph{classical} PDF stems from various approximations of $P_{\rm q}$ \cite{AkM}, which however do not alter this prefactor. This is the origin of the functional form of Eq.\ \Eq{Puv}.


\section{QUANTUM-MECHANICAL TIME}\label{qmtime}

In the previous section, we derived the PDF of a quantum diffusive process and found that, as in the standard case \cite{AkM}, it is of the form $P_{\rm q}\sim GG$, where $G$ is the propagator of the particle. The classical PDF $P$ can then be derived from $P_{\rm q}$. (It is too heuristic to regard the diffusion equation as the Wick-rotated version of the Schr\"odinger equation and the classical PDF $P$ as the Euclideanized version of the quantum propagator $G$.) We thus obtained the coordinate-dependent normalization \Eq{Puv}. This is not the only information one can extract from quantum mechanics. In fact, the time-space dependence of the propagator $G$, calculated from the path integral, fixes the natural quantum-mechanical time $T$, i.e., the time with respect to which the path integral yields the transition probability. Since the scaling ratio of this time with respect to spatial coordinates filters down to the classical level, this provides an independent identification of diffusion time and a check of the diffusion equations \Eq{tddcf} and \Eq{tddcf2} in the fixed-dimensionality case, where the relation between dispersion and diffusion time is given by Eq.~\Eq{ell2}. To this purpose, the calculation of the free-particle propagator $G_{\rm free}$ suffices. In what follows, the symbol $q$ always denotes canonical coordinates (one spatial dimension for simplicity), while for geometric coordinates we reserve the symbol $\vr(x)$.

\subsection{Weighted Lagrangian}

The quantum mechanics corresponding to the fractional scenarios with weighted Laplacians features derivatives $\cD$ in the Lagrangian. The free-particle action is
\be\label{actpart}
S=\int_{t'}^t\rmd t''\,v_0(t'')\, L\,,
\ee
where $L=\frac12 m (\cD_t q)^2$. The Green function as a path integral
\be\label{pain}
G\sim \sum_{\rm paths} e^{\rmi S}
\ee
was computed in \cite{frc5} for this system and it reads
\ba
G_{\rm free}(q,t;q',t') &=&\sqrt{\frac{v_0(t')}{v_0(t) v(q)v(q')}}\ \sqrt{\frac{\rmi m}{2\pi (t-t')}}\nonumber\\
&&\times \exp\left\{\frac{\rmi m[\sqrt{v_0(t)}q-\sqrt{v_0(t')}q']^2}{2(t-t')}\right\}.\nonumber\\\label{gdisc3}
\ea
The relative scaling between space and time coordinates thus roughly identifies the quantum-mechanical time
\be\label{qm1}
T \sim \frac{t}{v_0(t)}\,,
\ee
which, for a power-law measure, yields
\be\label{qmb}
T \propto t^{2-\b}\,,
\ee
in agreement with \Eq{ell22} in the case $\nu=1$. Equation \Eq{qm1} is not a rigorous definition stemming from Eq.\ \Eq{gdisc3}, where the time-space scaling ratio is quite entangled. This is the reason why we do not get a precise matching with Eq.\ \Eq{ell2}, which instead we obtain in the next example.

\subsection{Ordinary Lagrangian}

The standard-Laplacian scenario corresponds to a Lagrangian $L=\frac12 m \dot q^2$. For this case we report the full calculation, which is similar to the one in \cite{frc5}. Partitioning the time interval  $t-t'$ into $N$ infinitesimal parts, $t=t_N>t_{N-1}>\dots>t_1>t_0=t'$, the action is
\ba
S &=& \frac{m}{2}\sum_{n=0}^{N-1} \int_{t_n}^{t_{n+1}}\rmd t\,v_0(t) [\dot q(t)]^2\nonumber\\
&\approx&\frac{m}{2}\sum_{n=0}^{N-1} v_0(t_{n+1}) \frac{[q(t_{n+1})-q(t_{n})]^2}{t_{n+1}-t_n}\,,
\ea	
where in the second step one could replace $v_0(t_{n+1})$ with $v_0(t_{n})$ or the average $[v_0(t_{n+1})-v_0(t_n)]/2$ without changing the final result.
		
To sum over all trajectories and get \Eq{pain}, it is sufficient to integrate over all possible $q_n=q(t_n)$ (ordinary integration, without weights):
\ba
G_N&:=&K_N\int \rmd q_1\dots\rmd q_{N-1}\nonumber\\
&&\quad\times \exp\left[-\frac{m}{2\rmi\e}\sum_{n=0}^{N-1} v_0(t_{n+1})(q_{n+1}-q_n)^2\right],\nonumber
\ea
where $K_N$ is a constant and we chose a partition with identical segments $\e= t_{n+1}-t_n$, $N\epsilon=t-t'$. It is easy to show by iteration that
\ba
\hspace{-0.8cm} G_N(q_N,t_N;q_0,t_0) &=& K_N\prod_{n=1}^{N}\sqrt{\frac{2\pi\rmi\e}{m v_0(t_n)}}\sqrt{\frac{m}{2\pi\rmi}\frac{1}{T_N}}\nonumber\\
&&\quad\times\exp\left[-\frac{m}{2\rmi}\frac{(q_N-q_0)^2}{T_N}\right],
\ea
where
\be
T_N := \sum_{n=1}^{N}\frac{\e}{v_0(t_n)}\,.
\ee
The constant $K_N$ can be chosen so that $G_N(q_N,q_0)=\int\rmd q_M G_N(q_N,q_M)G_M(q_M,q_0)$, which implies $K_N=K_MK_{N-M}$. Noticing that this relation is satisfied by
\be
K_N=\prod_{n=1}^{N}\sqrt{\frac{m v_0(t_n)}{2\pi\rmi\e}}\,,
\ee
we thus have
\be
G_N(q_N,t_N;q_0,t_0)=\sqrt{\frac{m}{2\pi\rmi}\frac{1}{T_N}}\exp\left[-\frac{m}{2\rmi}\frac{(q_N-q_0)^2}{T_N}\right]\,.
\ee
In the double limit $N\to \infty$, $\e\to 0$, with $N\e=t-t'$ fixed, we finally get ($q_N=q$, $q_0=q'$)
\ba
\hspace{-0.8cm} G_{\rm free}(q,t;q',t')&:=&\lim_{N\to\infty}G_N(q,t;q',t')\nonumber\\
&=& \sqrt{\frac{m}{2\pi\rmi}\frac{1}{T}}\exp\left[-\frac{m}{2\rmi}\frac{(q-q')^2}{T}\right],\label{gdisc4}
\ea
where
\be\label{qm2}
T =\lim_{N\to\infty}T_N = \int_{t'}^t\frac{\rmd t''}{v_0(t'')}\,,
\ee
which coincides with Eq.\ \Eq{ell2} with $\nu=1$. For a power-law weight $v_0(t)\propto t^{\b-1}$, Eqs.\ \Eq{qm2} and \Eq{qm1} coincide with \Eq{qmb} up to a positive constant, but otherwise they are different.

\subsection{\texorpdfstring{$q$}{}-Lagrangian}

This case is obvious, and from the ordinary Green function it follows that $T=\vr(t)$. In the case of fixed dimensionality, $T\propto t^\b$.

\end{appendix}



\begin{thebibliography}{99}

\bibitem{AO}    S.\ Alexander and R.\ Orbach, \tia{Density of states on fractals: ``fractons''} \doinn{10.1051/jphyslet:019820043017062500}{J.\ Phys.\ Lett.}{43}{625}{1982}.
\bibitem{RT}    R.\ Rammal and G.\ Toulouse, \tia{Random walks on fractal structures and percolation clusters} \doinn{10.1051/jphyslet:0198300440101300}{J.\ Phys.\ Lett.}{44}{13}{1983}.
\bibitem{Wat85} H.\ Watanabe, \tia{Spectral dimension of a wire network} \doin{10.1088/0305-4470/18/14/030}{J.\ Phys.}{A}{18}{2807}{1985}.
\bibitem{HBA}   S.\ Havlin and D.\ Ben-Avraham, \tia{Diffusion in disordered media} \doinn{10.1080/00018738700101072}{Adv.\ Phys.}{36}{695}{1987}.
\bibitem{KiL}   J.\ Kigami and M.L.\ Lapidus, \tia{Weyl's problem for the spectral distribution of Laplacians on P.C.F.\ self-similar fractals} \doinn{10.1007/BF02097233}{Commun.\ Math.\ Phys.}{158}{93}{1993}.
\bibitem{bH}    D.\ ben-Avraham and S.\ Havlin, \book{Diffusion and Reactions in Fractals and Disordered Systems}{Cambridge University Press}{Cambridge}{England}{2000}.

\bibitem{fra4}  G.\ Calcagni, \tia{Discrete to continuum transition in multifractal spacetimes} \doin{10.1103/PhysRevD.84.061501}{Phys.\ Rev.}{D}{84}{061501(R)}{2011} [\arX{1106.0295}].
\bibitem{frc1}  G.\ Calcagni, \tia{Geometry of fractional spaces} \doinn{10.4310/ATMP.2012.v16.n2.a5}{Adv.\ Theor.\ Math.\ Phys.}{16}{549}{2012} [\arX{1106.5787}].
\bibitem{frc2}  G.\ Calcagni, \tia{Geometry and field theory in multi-fractional spacetime} \doij{10.1007/JHEP01(2012)065}{J.\ High Energy Phys.}{01}{065}{2012} [\arX{1107.5041}].
\bibitem{frc3}  G.\ Calcagni and G.\ Nardelli, \tia{Momentum transforms and Laplacians in fractional spaces} \doinn{10.4310/ATMP.2012.v16.n4.a5}{Adv.\ Theor.\ Math.\ Phys.}{16}{1315}{2012} [\arX{1202.5383}].
\bibitem{ACOS}  M.\ Arzano, G.\ Calcagni, D.\ Oriti, and M.\ Scalisi, \tia{Fractional and noncommutative spacetimes} \doin{10.1103/PhysRevD.84.125002}{Phys.\ Rev.}{D}{84}{125002}{2011} [\arX{1107.5308}].
\bibitem{fra6}  G.\ Calcagni, \tia{Diffusion in quantum geometry} \doin{10.1103/PhysRevD.86.044021}{Phys.\ Rev.}{D}{86}{044021}{2012} [\arX{1204.2550}].
\bibitem{frc4}  G.\ Calcagni, \tia{Diffusion in multiscale spacetimes} \doin{10.1103/PhysRevE.87.012123}{Phys.\ Rev.}{E}{87}{012123}{2013} [\arX{1205.5046}].
\bibitem{frc5}  G.\ Calcagni, G.\ Nardelli, and M.\ Scalisi, \tia{Quantum mechanics in fractional and other anomalous spacetimes} \doinn{10.1063/1.4757647}{J.\ Math.\ Phys.\ (N.Y.)}{53}{102110}{2012} [\arX{1207.4473}].
\bibitem{fra7}  G.\ Calcagni, \tia{Multi-fractional spacetimes, asymptotic safety and Ho\v{r}ava--Lifshitz gravity} \doin{10.1142/S0217751X13500929}{Int.\ J.\ Mod.\ Phys.}{A}{28}{1350092}{2013} [\arX{1209.4376}].
\bibitem{frc6}  G.\ Calcagni and G.\ Nardelli, \tia{Symmetries and propagator in multi-fractional scalar field theory} \doin{10.1103/PhysRevD.87.085008}{Phys.\ Rev.}{D}{87}{085008}{2013} [\arX{1210.2754}].
\bibitem{AIP}   G.\ Calcagni, \tia{Introduction to multifractional spacetimes} \doinn{10.1063/1.4756961}{AIP Conf.\ Proc.}{1483}{31}{2012} [\arX{1209.1110}].

%
\bibitem{fra1}  G.\ Calcagni, \tia{Fractal universe and quantum gravity} \doinn{10.1103/PhysRevLett.104.251301}{Phys.\ Rev.\ Lett.}{104}{251301}{2010} [\arX{0912.3142}].
\bibitem{fra2}  G.\ Calcagni, \tia{Quantum field theory, gravity and cosmology in a fractal universe} \doij{10.1007/JHEP03(2010)120}{J.\ High Energy Phys.}{03}{120}{2010} [\arX{1001.0571}].
\bibitem{fra3}  G.\ Calcagni, \tia{Gravity on a multifractal} \doin{10.1016/j.physletb.2011.01.063}{Phys.\ Lett.}{B}{697}{251}{2011} [\arX{1012.1244}].
\bibitem{Car09} S Carlip, \tia{Spontaneous dimensional reduction in short-distance quantum gravity?} \doinn{10.1063/1.3284402}{AIP Conf.\ Proc.}{1196}{72}{2009} [\arX{0909.3329}].
\bibitem{SVW2}  T.P.\ Sotiriou, M.\ Visser, and S.\ Weinfurtner, \tia{From dispersion relations to spectral dimension ---and back again}
  \doin{10.1103/PhysRevD.84.104018}{Phys.\ Rev.}{D}{84}{104018}{2011} [\arX{1105.6098}].
\bibitem{CES}   G.\ Calcagni, A.\ Eichhorn, and F.\ Saueressig, \doin{10.1103/PhysRevD.87.124028}{Phys.\ Rev.}{D}{87}{124028}{2013} [\arX{1304.7247}].
\bibitem{Pad98} T.\ Padmanabhan, \tia{Quantum structure of space-time and black hole entropy} \doinn{10.1103/PhysRevLett.81.4297}{Phys.\ Rev.\ Lett.}{81}{4297}{1998} [\arX{hep-th/9801015}].
\bibitem{Pad99} T.\ Padmanabhan, \tia{Event horizon: magnifying glass for Planck length physics} \doin{10.1103/PhysRevD.59.124012}{Phys.\ Rev.}{D}{59}{124012}{1999} [\arX{hep-th/9801138}].
\bibitem{AC1}   M.\ Arzano and G.\ Calcagni, \tia{Black-hole entropy and minimal diffusion} \doin{10.1103/PhysRevD.88.084017}{Phys.\ Rev.}{D}{88}{084017}{2013} [\arX{1307.6122}].
\bibitem{Sok12} I.M.\ Sokolov, \tia{Models of anomalous diffusion in crowded environments} \doinn{10.1039/C2SM25701G}{Soft Matter}{8}{9043}{2012}.

\bibitem{Fic55} A.\ Fick, \tia{\"Uber Diffusion} Pogg.\ Ann.\ Phys.\ Chem.\ {\bf 170}, 59 (1855).
\bibitem{Ein05} A.\ Einstein, \tia{\"Uber die von der molekularkinetischen Theorie der W\"arme geforderte Bewegung von in ruhenden Fl\"ussigkeiten suspendierten Teilchen}
  \doinn{10.1002/andp.19053220806}{Ann.\ Phys.\ (Berlin)}{322}{549}{1905}.
\bibitem{Smo06} M.\ von Smoluchowski, \tia{Zur kinetischen Theorie der Brownschen Molekularbewegung und der  Suspensionen} \doinn{10.1002/andp.19063261405}{Ann.\ Phys.\ (Berlin)}{326}{756}{1906}.
\bibitem{KTH}   R.\ Kubo, M.\ Toda, and N.\ Hashitsume, \books{Statistical Physics II---Nonequilibrium Statistical Mechanics}{Springer-Verlag}{Berlin}{1985}. 
\bibitem{Zwa01} R.\ Zwanzig, \books{Nonequilibrium Statistical Mechanics}{Oxford University Press}{Oxford}{2001}.

\bibitem{WaL}   K.G.\ Wang and C.W.\ Lung, \tia{Long-time correlation effects and fractal Brownian motion} \doin{10.1016/0375-9601(90)90175-N}{Phys.\ Lett.}{A}{151}{119}{1990}.
\bibitem{Wan92} K.G.\ Wang, \tia{Long-time-correlation effects and biased anomalous diffusion} \doin{10.1103/PhysRevA.45.833}{Phys.\ Rev.}{A}{45}{833}{1992}.
\bibitem{MaW} J.\ Masoliver and K.G.\ Wang, \tia{Free inertial processes driven by Gaussian noise: probability distributions, anomalous diffusion, and fractal behavior}
  \doin{10.1103/PhysRevE.51.2987}{Phys.\ Rev.}{E}{51}{2987}{1995}.
\bibitem{WaT}   K.G.\ Wang and M.\ Tokuyama, \tia{Nonequilibrium statistical description of anomalous diffusion} \doinn{10.1016/S0378-4371(98)00644-X}{Physica (Amsterdam)}{265A}{341}{1999}.
\bibitem{Lut01} E.\ Lutz, \tia{Fractional Langevin equation} \doin{10.1103/PhysRevE.64.051106}{Phys.\ Rev.}{E}{64}{051106}{2001}.
\bibitem{LiM02} S.C.\ Lim and S.V.\ Muniandy, \tia{Self-similar Gaussian processes for modeling anomalous diffusion} \doin{10.1103/PhysRevE.66.021114}{Phys.\ Rev.}{E}{66}{021114}{2002}.
\bibitem{MeK04} E.\ Metzler and J.\ Klafter, \tia{The restaurant at the end of the random walk: recent developments in the description of anomalous transport by fractional dynamics}
  \doin{10.1088/0305-4470/37/31/R01}{J.\ Phys.}{A}{37}{R161}{2004}.
\bibitem{JeM}   J.-H.\ Jeon and R.\ Metzler, \tia{Fractional Brownian motion and motion governed by the fractional Langevin equation in confined geometries}
  \doin{10.1103/PhysRevE.81.021103}{Phys.\ Rev.}{E}{81}{021103}{2010} [\arX{1001.0681}].

%
\bibitem{MeK00} R.\ Metzler and J.\ Klafter, \tia{The random walk's guide to anomalous diffusion: a fractional dynamics approach} \doinn{10.1016/S0370-1573(00)00070-3}{Phys.\ Rep.}{339}{1}{2000}.
\bibitem{Zas02} G.M.\ Zaslavsky, \tia{Chaos, fractional kinetics, and anomalous transport} \doinn{10.1016/S0370-1573(02)00331-9}{Phys.\ Rep.}{371}{461}{2002}.

\bibitem{Avr00} I.G.\ Avramidi, \books{Heat Kernel and Quantum Gravity}{Springer-Verlag}{Berlin}{2000}. 
\bibitem{Kir01} K.\ Kirsten, \book{Spectral Functions in Mathematics and Physics}{Chapman \& Hall/CRC}{Boca Raton}{U.S.A.}{2001}.
\bibitem{Vas03} D.V.\ Vassilevich, \tia{Heat kernel expansion: user's manual} \doinn{10.1016/j.physrep.2003.09.002}{Phys.\ Rep.}{388}{279}{2003} [\arX{hep-th/0306138}].

\bibitem{RYS}   F.-Y.\ Ren, Z.-G.\ Yu, and F.\ Su, \tia{Fractional integral associated to the self-similar set or the generalized self-similar set and its physical interpretation} \doin{10.1016/0375-9601(96)00418-5}{Phys.\ Lett.}{A}{219}{59}{1996}.
\bibitem{RLWQ}  F.-Y.\ Ren, J.-R.\ Liang, X.-T.\ Wang, and W.-Y.\ Qiu, \tia{Integrals and derivatives on net fractals} \doinn{10.1016/S0960-0779(02)00211-4}{Chaos Solitons Fractals}{16}{107}{2003}.
\bibitem{NLM}   R.R.\ Nigmatullin and A.\ Le M\'ehaut\'e, \tia{Is there geometrical/physical meaning of the fractional integral with complex exponent?} \doinn{10.1016/j.jnoncrysol.2005.05.035}{J.\ Non-Cryst.\ Solids}{351}{2888}{2005}.

%
\bibitem{Akk12} E.\ Akkermans, \tia{Statistical mechanics and quantum fields on fractals} in \emph{Fractal Geometry and Dynamical Systems in Pure and Applied Mathematics I: Fractals in Pure Mathematics}, edited by D.\ Carfi, M.L.\ Lapidus, E.P.J.\ Pearse, and M.\ van Frankenhuijsen (AMS, Providence, U.S.A., 2013) [\arX{1210.6763}].
\bibitem{Akk2}  E.\ Akkermans, G.V.\ Dunne, and A.\ Teplyaev, \tia{Thermodynamics of photons on fractals} \doinn{10.1103/PhysRevLett.105.230407}{Phys.\ Rev.\ Lett.}{105}{230407}{2010} [\arX{1010.1148}].

\bibitem{SmS1}  A.\ Smailagic and E.\ Spallucci, \tia{Feynman path integral on the noncommutative plane} \doin{10.1088/0305-4470/36/33/101}{J.\ Phys.}{A}{36}{L467}{2003}
 [\arX{hep-th/0307217}].
\bibitem{SmS2}  A.\ Smailagic and E.\ Spallucci, \tia{UV divergence free QFT on noncommutative plane} \doin{10.1088/0305-4470/36/39/103}{J.\ Phys.}{A}{36}{L517}{2003} [\arX{hep-th/0308193}].
\bibitem{SmS3}  A.\ Smailagic and E.\ Spallucci, \tia{Lorentz invariance, unitarity in UV-finite of QFT on noncommutative spacetime} \doin{10.1088/0305-4470/37/28/008}{J.\ Phys.}{A}{37}{7169}{2004} [\arX{hep-th/0406174}].
\bibitem{SSM}   E.\ Spallucci, A.\ Smailagic, and P.\ Nicolini, \tia{Trace anomaly in quantum spacetime manifold} \doin{10.1103/PhysRevD.73.084004}{Phys.\ Rev.}{D}{73}{084004}{2006} [\arX{hep-th/0604094}].
\bibitem{Rin09} M.\ Rinaldi, \tia{A new approach to non-commutative inflation} \doinn{10.1088/0264-9381/28/10/105022}{Classical Quantum Gravity}{28}{105022}{2011} [\arX{0908.1949}].
\bibitem{MoN}   L.\ Modesto and P.\ Nicolini, \tia{Spectral dimension of a quantum universe} \doin{10.1103/PhysRevD.81.104040}{Phys.\ Rev.}{D}{81}{104040}{2010} [\arX{0912.0220}].
\bibitem{NiN}   P.\ Nicolini and B.\ Niedner, \tia{Hausdorff dimension of a particle path in a quantum manifold} \doin{10.1103/PhysRevD.83.024017}{Phys.\ Rev.}{D}{83}{024017}{2011}
  [\arX{1009.3267}].

\bibitem{BaA}   J.A.\ Barnes and D.W.\ Allan, \tia{A statistical model of flicker noise} \doinn{10.1109/PROC.1966.4630}{Proc.\ IEEE}{54}{176}{1966}. 
\bibitem{MaV}   B.B.\ Mandelbrot and J.W.\ Van Ness, \tia{Fractional Brownian motions, fractional noises and applications} \doinn{10.1137/1010093}{SIAM Rev.}{10}{422}{1968}.

%
\bibitem{GR}    I.S.\ Gradshteyn and I.M.\ Ryzhik, \books{Table of Integrals, Series, and Products}{Academic Press}{London}{2000}.
\bibitem{AkM}   E.\ Akkermans and G.\ Montambaux, \book{Mesoscopic Physics of Electrons and Photons}{Cambridge University Press}{Cambridge}{England}{2011}.
\end{thebibliography}
\end{document}